\documentclass[nofootinbib,amsmath,amssymb,amsfonts,aps,prd,eqsecnum,superscriptaddress,reprint]{revtex4-1}

\pdfoutput=1
\usepackage{hyperref}
\usepackage{amsmath}
\usepackage{amssymb}
\usepackage{mathrsfs}
\usepackage{xcolor,graphicx}
\usepackage[caption=false]{subfig}
\usepackage[toc,page]{appendix}
\usepackage[section]{placeins}
\usepackage{breqn}
\usepackage[T1]{fontenc}
\usepackage[normalem]{ulem}
\usepackage{soul}

\def\fref{Fig.~\ref}
\def\sref{\S~\ref}
\def\eref{Eq.~\ref}

\makeatletter
\let\cat@comma@active\@empty
\makeatother

\begin{document}

\title{Multimessenger Asteroseismology of Core-Collapse Supernovae}

\author{John Ryan Westernacher-Schneider}
\email{jwestern@email.arizona.edu}
\affiliation{Department of Astronomy/Steward Observatory, The University of Arizona, 933 N. Cherry Ave, Tucson, AZ 85721, USA}
\affiliation{Department of Physics, University of Guelph, Guelph, Ontario N1G 2W1, Canada}
\affiliation{Perimeter Institute for Theoretical Physics, 31 Caroline Street North, Waterloo, Ontario N2L 2Y5, Canada}
\affiliation{DARK, Niels Bohr Institute, University of Copenhagen,   Lyngbyvej 2, 2100, Copenhagen, Denmark}

\author{Evan O'Connor}
\email{evan.oconnor@astro.su.se}
\affiliation{Department of Astronomy and The Oskar Klein Centre, Stockholm University, AlbaNova, 109 61, Stockholm, Sweden}
  
\author{Erin O'Sullivan}
\email{erin.osullivan@fysik.su.se}
\affiliation{The Oskar Klein Centre and the Department of Physics,  Stockholm University, AlbaNova, 109 61, Stockholm, Sweden}

\author{Irene Tamborra}
\email{tamborra@nbi.ku.dk}
\affiliation{Niels Bohr International Academy, Niels Bohr Institute, University of Copenhagen, 2100, Copenhagen, Denmark}
\affiliation{DARK, Niels Bohr Institute, University of Copenhagen, Lyngbyvej 2, 2100, Copenhagen, Denmark}

\author{Meng-Ru Wu}
\email{mwu@gate.sinica.edu.tw}
\affiliation{Institute of Physics,  Academia Sinica,  Taipei, 11529 Taiwan}
\affiliation{Institute of Astronomy and Astrophysics,  Academia Sinica, 10617  Taipei, Taiwan}
\affiliation{Niels Bohr International Academy, Niels Bohr Institute, University of Copenhagen, 2100, Copenhagen, Denmark}

\author{Sean M. Couch}
\email{couch@pa.msu.edu}
\affiliation{Department of Physics and Astronomy, Michigan State University, East Lansing, MI 48824, USA}
\affiliation{Department of Computational Mathematics, Science, and Engineering, Michigan State University, East Lansing, MI 48824, USA}
\affiliation{National Superconducting Cyclotron Laboratory, Michigan State University, East Lansing, MI 48824, USA}
\affiliation{Joint Institute for Nuclear Astrophysics-Center for the Evolution of the Elements, Michigan State University, East Lansing, MI 48824, USA}

\author{Felix Malmenbeck}
\email{felixm@kth.se}
\affiliation{KTH Royal Institute of Technology, School of Engineering Sciences (SCI), Physics. Roslagstullsbacken 21, SE-10691 Stockholm, Sweden}


\begin{abstract}
We investigate correlated gravitational wave and neutrino signals from rotating core-collapse supernovae with simulations. Using an improved mode identification procedure based on mode function matching, we show that a linear quadrupolar mode of the core produces a dual imprint on gravitational waves and neutrinos in the early post-bounce phase of the supernova. The angular harmonics of the neutrino emission are consistent with the mode energy around the neutrinospheres, which points to a mechanism for the imprint on neutrinos. Thus, neutrinos carry information about the mode amplitude in the outer region of the core, whereas gravitational waves probe deeper in. We also find that the best-fit mode function has a frequency bounded above by $\sim 420$ Hz, and yet the mode's frequency in our simulations is $\sim 15\%$ higher, due to the use of Newtonian hydrodynamics and a widely used pseudo-Newtonian gravity approximation. This overestimation is particularly important for the analysis of gravitational wave detectability and asteroseismology, pointing to limitations of pseudo-Newtonian approaches for these purposes, possibly even resulting in excitation of incorrect modes. In addition, mode frequency matching (as opposed to mode function matching) could be resulting in mode misidentification in recent work. Lastly, we evaluate the prospects of a multimessenger detection of the mode using current technology. The detection of the imprint on neutrinos is most challenging, with a maximum detection distance of $\sim\!1$\,kpc using the IceCube Neutrino Observatory. The maximum distance for detecting the complementary gravitational wave imprint is $\sim\!5$\,kpc using Advanced LIGO at design sensitivity.
\end{abstract}

\maketitle

\section{Introduction}

The collapse and bounce of the iron cores of massive ($M\gtrsim 10 M_{\odot}$) stars and the possible ensuing explosion are expected to produce detectable gravitational waves (GWs) and neutrinos if they occur within or nearby our galaxy. Indeed, neutrinos have already been detected from such an event, namely SN1987A~\cite{hirata1987observation,bionta1991observation}. Core collapse events with a successful explosion are called core-collapse supernovae (CCSNe). The electron-degenerate iron core collapses once it exceeds its effective Chandrasekhar mass limit, and is halted once the core reaches nuclear densities $\rho \sim \mathrm{few}\times 10^{14}$\,g\,cm$^{-3}$ and its equation of state stiffens. The core overshoots its equilibrium radius, resulting in an overpressure, and bounces outwards again. This imparts momentum to the supersonically infalling stellar material, causing a powerful outward shockwave. Whether and how this shockwave and the subsequent dynamics result in a successful explosion is a central theme of research in this area, see eg. the recent reviews~\cite{muller2016core,couch2017mechanism} and references therein.

In the event of a successful explosion, photons will also be detectable.  In contrast with photons, which are heavily reprocessed before freely streaming to an observer, the intervening stellar material between the core and an observer are transparent to GWs. The star is also largely transparent to neutrinos, except the region within $\sim50$\,km of the centre of the proto-neutron star (PNS) where neutrino-matter interactions are still strong. Neutrinos and GWs therefore offer direct probes of the central engine of a CCSN~\cite{Janka:2017vlw,Muller:2019upo,Mirizzi:2015eza,Muller:2017vuu,Aasi:2013wya}.

Gravitational waves in CCSNe arise from coherent matter accelerations. One of the strongest sources of GWs in CCSNe is from strongly rotating core collapse.  In this scenario, the collapsing rotating core has an accelerating quadrupole moment, and therefore generates gravitational waves.   At core bounce, the newly formed rotating PNS emits a distinct GW pattern. Following core bounce and the stagnation of the CCSN shock, the growth of turbulence at $\gtrsim$\,100-150\,ms after bounce can also lead to excitations of the PNS and the production of GWs. These are often emitted by characteristic modes of the PNS ($f_{\mathrm{GW}} \gtrsim 500$\,Hz). GW emission can occur at lower frequencies as well ($f_{\mathrm{GW}} \sim 100-200$\,Hz), due to matter motions further out where the dynamic timescale is longer.  In this work, we focus on the time interval $0\,\mathrm{ms} \lesssim t_{\mathrm{pb}} \lesssim 150\,\mathrm{ms}$.  In addition to low-frequency GWs from interactions between the prompt convection and the shock \cite{mueller2013gw}, GWs may also be expected from the PNS as it settles down after the very dynamic bounce phase. Contrasting against the late-time signal, the PNS radius is larger and the mass is lower, so one may expect lower frequency GWs. For a more in-depth review of GWs from CCSNe and the different emission regimes we refer the reader to a recent review \cite{kotate:SNhandbook}. Interestingly, correlated frequencies between GWs and neutrino luminosities have been observed in simulations within tens of milliseconds after core bounce in~\cite{ott2012correlated}. Similar correlations at later times have been reported in~\cite{Tamborra:2013laa,Tamborra:2014hga,kuroda2017correlated,Takiwaki:2017tpe,Walk:2019ier,Walk:2018gaw,Andresen:2018aom}, purportedly due to the growth of the standing accretion-shock instability (SASI). These observations point towards a wealth of opportunity to probe specific aspects of the central dynamics occurring at different times, from tens of ms to several seconds and beyond.

Asteroseismology is the study of the interior structure of stars inferred from observations of its seismic oscillations. There have been recent theoretical efforts to use GWs to do the same with CCSNe, so-called \emph{gravitational wave asteroseismology}~\cite{andersson1998towards,fuller2015supernova,torres2017towards, morozova2018gravitational,torres2018towards}. These efforts involve identifying the modes responsible for GW emission in simulations. The main strategy is to use data from numerical simulations as input for a perturbative mode calculation, where the simulated data serves as a background solution. The key point to note is that there is a separation of scale between the period of the modes of interest and the time scale over which the post-bounce CCSN background changes significantly. For example, towards the pessimistic end, a $200$ Hz mode has a period of $5$ ms, whereas the CCSN background changes over a timescale of several tens of ms. Therefore one expects to be able to treat the CCSN background as stationary for the purposes of a perturbative calculation at any instant of time. In~\cite{fuller2015supernova} this was done using a perturbative Newtonian hydrodynamic scheme in an effort to generate a qualitative understanding of the GW emission due to the oscillations of a rotating PNS that were excited at bounce. Subsequently,~\cite{torres2017towards} presented a similar effort using perturbative hydrodynamic calculations in the relativistic Cowling approximation. Shortly thereafter, and during the course of this work,~\cite{morozova2018gravitational} partially relaxed the Cowling approximation by allowing the lapse to vary, governed by the Poisson equation. They claimed an improved coincidence between their perturbative mode frequencies and certain emission features in the GW spectrograms from simulations. The Cowling approximation was then relaxed even further in~\cite{torres2018towards}, where the conformal factor of the spatial metric was allowed to vary as well, leaving only the shift vector fixed.

All of these studies, however, attempt to identify specific modes of oscillation of the system primarily by coincidence between perturbative mode frequencies and peaks in GW spectra, across time. This is potentially problematic for a number of reasons. Firstly, the approximations used in the perturbative calculations introduce errors in mode frequencies that can be quite significant, e.g.,~tens of percent in the case of lower order modes in the Cowling approximation. Secondly, any partial relaxation of the Cowling approximation presents difficulties with the interpretation of results, since the resulting perturbative scheme neglects some terms at a given order but not others, and thus is not under control; one cannot argue \emph{a priori} that the neglected terms are smaller than those included, and so the regime of applicability requires independent investigation. Furthermore, the perturbative schemes applied in~\cite{torres2017towards,morozova2018gravitational,torres2018towards} are not consistent linearizations of the equations being solved in the simulations, albeit they are inconsistent in different ways due to different 
perturbative schemes and simulation methodologies. Thirdly, the approximations used in the simulations themselves introduce their own frequency errors. For example, often hydrodynamics is treated as Newtonian and gravity treated in a pseudo-Newtonian manner in CCSN simulations by modifying the potential to mimic relativistic effects, as in~\cite{marek2006exploring,morozova2018gravitational,o2018two,pan2018equation}. Since the mode population in the vicinity of a given frequency bin in a GW spectrum can be rather dense (neighboring mode frequencies differing by $\sim 5-10\%$), and often the temporal evolution of neighboring mode frequencies are approximately related by a scalar multiple ($\sim 1.05-1.10$), all of these above mentioned sources of error serve to lower the significance of any given observed coincidence between perturbative and simulated mode frequencies. Indeed, the purported identification of a $g$-mode in~\cite{pan2018equation} required a \emph{post hoc} modification of its frequency formula when matching to GW spectrograms from simulations, which was speculated to be due to the use of Newtonian hydrodynamics and pseudo-Newtonian gravity in the simulations.

We opt to take a different approach. Rather than looking for coincidence in mode frequencies, we look for coincidence in \emph{mode functions}. This means comparing mode functions, obtained from perturbative calculations, with the velocity data from simulations, which are post-processed using spectral filters and vector spherical harmonic decompositions. Mode function matching and mode frequency matching may not agree unless the perturbative scheme applied is a consistent linearization of the equations being solved in the simulations. The perturbative schemes in ~\cite{torres2017towards,morozova2018gravitational,torres2018towards} are not consistent linearizations of the simulations they are applied to, so one expects mode function matching can give different results than mode frequency matching. We find that if a mode has an adequate excitation, its matching with candidate perturbative mode functions produces an unambiguous best-fit; see~\cite{westernacher2018turbulence} for an exhaustive demonstration. Since this strategy does not use frequency-matching, we also discover the frequencies observed in our simulations are overestimated, with the true values being about $15$\% lower. This illustrates the power of matching in mode functions rather than mode frequencies, and bears out our concerns with focusing only on mode frequency coincidence as in~\cite{torres2017towards,morozova2018gravitational,torres2018towards}. These results were reported previously in~\cite{westernacher2018turbulence} with an erroneously large frequency discrepancy of $\sim 40\%$, which we correct in this work to $\sim 15\%$.

Since the demonstration of mode function matching in~\cite{westernacher2018turbulence}, the partially-relaxed Cowling approximation of~\cite{morozova2018gravitational} and mode identification via frequency coincidence was used again in~\cite{radice2019characterizing}. In~\cite{torres2019universal}, a reclassification of previously misidentified modes from~\cite{torres2018towards} was proposed, of the sort one would expect based on our concerns outlined above ($\pm 1$ miscounts of radial nodes $n$). Very recently, in~\cite{sotani2019dependence} the frequency spectrum was computed on a fully general relativistic simulated background CCSN using the relativistic Cowling approximation. Mode frequency matching was again used in an attempt to identify the active modes in the simulation, even though the Cowling approximation will be systematically overestimating the frequencies of the modes present in the general relativistic simulation. Correlations between neutrino and gravitational wave emission and properties of the central CCSN engine (over longer timescales than we study here) were then explored in~\cite{vartanyan2019temporal}, where the mode frequency matching methods of~\cite{morozova2018gravitational} were used again.

We use the perturbative scheme in the relativistic Cowling approximation from~\cite{torres2017towards}, which applies only to spherical systems. Therefore we only apply it to a non-rotating model in order to identify modes of oscillation that are excited at bounce and ring for $\sim 10-100$\,ms. This identification serves to label the corresponding modes in the rotating models~\cite{friedman2013rotating} whose mode functions deform continuously with increasing rotation, picking up a mixed character in angular harmonics. We also simulate a sequence of rotating models with progressively larger pre-collapse rotations of $\Omega_c\!=\!\lbrace 0.0, 0.5, 1.0, 1.5, 2.0, 2.5 \rbrace$\,rad\,s$^{-1}$. In order to follow the modes along this sequence, we take inspiration from the works of~\cite{font2001axisymmetric,dimmelmeier2006non,gaertig2008oscillations}. In~\cite{font2001axisymmetric}, knowledge of the modes of the non-rotating star were combined with continuity in frequency to follow modes across such a sequence, whereas in~\cite{dimmelmeier2006non,gaertig2008oscillations} they used continuity in the deformation of mode functions with varying rotation. 

Continuity in mode function is more powerful than continuity in frequency, since separate modes can have very similar frequencies and thus would be degenerate in a frequency continuity analysis. Thus we chiefly use mode function continuity to follow modes along our sequence of rotating models. We follow a particular quadrupolar mode successfully to the $\Omega_c\!=\! 1.0$\,rad\,s$^{-1}$ model that we focus on. When following the mode to larger rotations we find ambiguities, so we make the more conservative conclusion than in~\cite{westernacher2018turbulence} that we lose track of the mode beyond the $\Omega_c\!=\! 1.0$\,rad\,s$^{-1}$ model.

\emph{Our chief result is the implication of a linear quadrupolar mode in the }$\Omega_c\!=\! 1.0$\,rad\,s$^{-1}$\,\emph{model as imprinting on the GWs and neutrino emission, and the mechanism of this dual imprint.} We also demonstrate our improved mode identification via mode function matching, some variant of which could also be used to study, for example, pulsations of binary neutron star post-merger remnants or accretion-induced collapse of white dwarfs.

In contrast with~\cite{ott2012correlated} where the neutrino treatment did not supply information about the emission pattern on the sky, our treatment does allow this. We relate the dominant angular harmonics of the (spectrally-filtered) emission to the dominant energy harmonics in the $l=2$ mode function in the vicinity of the neutrinospheres. The causal explanation for the oscillations in the neutrino emission properties is that the $l=2$ mode of the PNS, which in the rotating $\Omega_c = 1.0$ rad$\,$s$^{-1}$ model has a mixed character in $l$, is producing $l=2$ and $l=0$ variations of the neutrinospheres. Since the neutrinospheres are roughly the boundary between trapped and free-streaming neutrinos, this means that the region producing free-streaming neutrinos is undergoing variations with an angular structure in accordance with the activity of the mode in the vicinity of the neutrinospheres, at $r\sim60-80$ km. This causes the oscillations in neutrino signal registered by an observer far away. 

We therefore find that detailed asteroseismology of CCSN is possible in principle with joint detection of GWs and neutrinos, where the neutrinos supply information from the neutrinosphere region $r \sim 60-80$ km, and the GWs supply information from deeper in. However, the 3 $\sigma$ discovery potential with present-day neutrino detectors like the IceCube Neutrino Observatory~\cite{Achterberg:2006md,Abbasi:2011ss,Kopke:2017req} is $\sim\!1$\,kpc for  the signatures and models investigated in this work. We expect the complementary GW signal to be observable in Advanced LIGO (assuming design sensitivity) for supernovae located within $\sim\!5$\,kpc.

This paper is organized as follows. In \sref{sec:FLASHdescribe} we introduce our CCSN models and describe our simulations. In \sref{sec:sims} we present the multimessenger signals and predicted IceCube neutrino rates from our simulations. \sref{sec:pertschemes} gives a brief description of the perturbative schemes of~\cite{torres2017towards,morozova2018gravitational}. \sref{sec:modeextract} describes our mode function matching procedure first reported in~\cite{westernacher2018turbulence}. The mode analysis and mechanism of dual imprint are presented in \sref{sec:dualimprint}, and multimessenger detection prospects are presented in \sref{sec:detect}. Mode tests of our perturbative schemes and our simulation code are performed on a stable hydrostatic star (Tolman-Oppenheimer-Volkoff (TOV) star) in Appendices \sref{sec:CCSN_test_pert_schemes} and \sref{sec:FLASHTOV_test}, respectively. When applied to the TOV star (which we note is more compact than a PNS and thus is a more demanding application), we find that the scheme of~\cite{morozova2018gravitational}, which we dub a \emph{partially-relaxed Cowling approximation} or simply a  \emph{partial Cowling approximation}, is significantly less accurate than the Cowling approximation itself for fundamental mode frequencies, and even fails to reproduce the correct radial order of high-order modes. The factors which affect neutrino detectability are investigated in Appendix \sref{sec:modelneutrino} using toy models. In Appendix \sref{sec:modedurability} we explore the sensitivity of the mode identification to different choices of boundary conditions, as well as compare the application of the Cowling approximation and the partial Cowling approximation of~\cite{morozova2018gravitational} to the CCSN system. In Appendix \sref{sec:kernels} we provide a sampling of the spectral filter kernels we use. We conclude in \sref{ch:CCSNconc}.

%
%

\section{Models and Methodology}\label{ch:CCSNimp}

In this section, we describe our numerical simulations, initial conditions, and the resulting neutrino and GW signals. 

\subsection{Numerical Simulations and Progenitor Models} \label{sec:FLASHdescribe}
To simulate rotating CCSNe,  we use the massively-parallel \texttt{FLASH} simulation framework \cite{fryxell2000flash,dubey2009extensible}.  \texttt{FLASH} offers tools for simulating compressible hydrodynamics.  These tools have been extended in order to simulate CCSNe, including support for nuclear equations of state, grid-based energy-dependent neutrino transport, and an effective general relativistic potential \cite{couch13,o2018two}. For the hydrodynamics, we use a fifth order WENO (weighted essentially non-oscillatory) reconstruction, an HLLC Reimann solver (but revert to a more diffusive HLLE solver in the presence of shocks), and a second order Runge-Kutta time integrator using the method of lines.  Details of this hydrodynamic solver will be presented in \cite{couch:2019}. Our computational grid is cylindrical.  The resolution in the core (and out to $\sim$\,80\,km) is $\sim$\,195\,m, and outside $\sim$\,80\,km we enforce refinement such that $\Delta x/r < 0^\circ.33$.  Of particular interest to this work is the treatment of gravity and neutrinos, which we describe in some detail below.

The gravity and hydrodynamic treatments are Newtonian, however an effective general-relativistic potential obtained through phenomenological considerations and tested in CCSN evolutions has been introduced in~\cite{Keil1997,rampp2002radiation,marek2006exploring} and implemented in our \texttt{FLASH} simulations in~\cite{couch13,o2018two}. The effective potential we use is a recasting of the monopole term of a multipole decomposition of the Newtonian gravitational potential. It is designed to recover the structure of relativistic stars in spherical symmetry. We retain the additional, non-spherical, Newtonian multipole moments for $1 \leq \ell \leq 16$ using the multipole solver of \cite{couch13b}. Since we do not solve for the gravitational metric, GWs are not actually present in the computational domain. We instead extract the GW signal using the quadrupole formula~\cite{thorne1980multipole,oconnorcouch18,pajkos2019gw}. For axisymmetric simulations, the only non-zero GW polarization is the $h_+$ polarization. This signal peaks for an observer situated in the equatorial plane and is vanishing for an observer along the axis of symmetry.  It is worth commenting on the impact of using the effective general relativistic potential to model the gravitational field and the use of the quadrupole formula to extract the GW signal. The use of the quadrupole formula has been validated in the context of rotating stellar core collapse and shown to give excellent results when compared to far field extraction techniques \cite{reisswig2011gravitational}. The effective potential, as we shall explore more in this paper, impacts the frequency spectrum of the emitted GWs. The dominant cause of this difference is that the underlying Newtonian hydrodynamics is not subject to the general relativistic kinematics \cite{mueller2013gw}, in particular the use of pseudo-Newtonian gravity and the absence of a lapse function in the hydrodynamic fluxes.

In CCSNe, neutrinos are present in both equilibrium and non-equilibrium states.  Simulating neutrinos requires a sophisticated treatment that accurately captures both of these regimes, and most importantly, the transition region between them.  A full solution, i.e. solving the energy-, species-, and angle-dependent Boltzmann equation, is limited by the large dimensionality of phase space and too computationally expensive to solve without some approximations in the methods or sacrifices in the resolution (see eg.~\cite{nagakura2018simulations,harada2019neutrino} for the latter). Many different approximate treatments have been employed in the literature. We choose to keep the energy dependence (18 energy groups) and approximate the angular dependence of the neutrino field by evolving only moments (in our case, the zeroth and first moments) of the Boltzmann equation \cite{shibata:11,cardall:13,o2015open}.  This method requires a closure. We choose the M1 closure where we analytically prescribe the second moment of the neutrino radiation field, the Eddington tensor. Unlike other approximations such as leakage or ray-by-ray, our method locally captures the neutrino emission (and absorption) and then transports the neutrinos directly on the multidimensional computational grid. This provides a great advantage compared to previous
work which studied the correlations between neutrinos and GWs from rotating core collapse in the past~\cite{ott2012correlated}, in particular because it provides directional emission information. Full details of our implementation in \texttt{FLASH} can be found in \cite{o2018two}.

With this computational setup we evolve a $20\, M_\odot$ zero-age main sequence mass presupernova progenitor model from the widely used stellar evolution calculations of Woosley \& Heger~\cite{woosley2007nucleosynthesis}.  We utilize the SFHo equation of state~\cite{steiner2013core,oconnor:10}, which is a modern tabulated nuclear equation of state compatible with the constraints available from eg. astrophysical observations of neutron stars. Neutrino microphysics is incorporated via \texttt{NuLib}~\cite{o2015open} and is chosen to match the setup of \cite{oconnoretal:18}.\footnote{The difference between the \texttt{FLASH} simulations here and those of~\cite{oconnoretal:18} are (a) our simulations are 2D and include rotation, (b) we include neutrino-electron inelastic scattering, thereby allowing an accurate evolution in \texttt{FLASH} during the collapse phase, and (d) we use a new hydrodynamic solver as discussed above and in~\cite{couch:2019}.} The $20\, M_\odot$ model was obtained from spherically symmetric stellar evolution calculations without rotation. 

We study a sequence of rotating models, and therefore we initialize this model with a pre-collapse rotation profile prescribed by hand. The initial rotation law imposed is taken to be,
\begin{eqnarray}
\Omega(r) = \frac{\Omega_{c}}{ 1 + \left( \frac{r}{A}\right)^2 }, \label{eq:rotlaw}
\end{eqnarray}
where $r = \sqrt{\varrho^2 + z^2}$ is the radial distance from the center, $\varrho$ is the cylindrical radius, and $z$ is the axial position.  For values of $r < A$, this gives roughly constant angular velocity of $\Omega_c$, i.e. solid body rotation.   For $r$ much greater than $A$ the star is described by constant specific angular momentum. For all the simulations presented here we adopt $A=800$\,km. The angular velocity of the fluid is taken to be $v_\phi(\varrho, z) = \varrho \Omega(r)$. 

\subsection{Signals in Advanced LIGO and IceCube}
\label{sec:sims}

\begin{figure}
\centering
\includegraphics[width=0.48\textwidth]{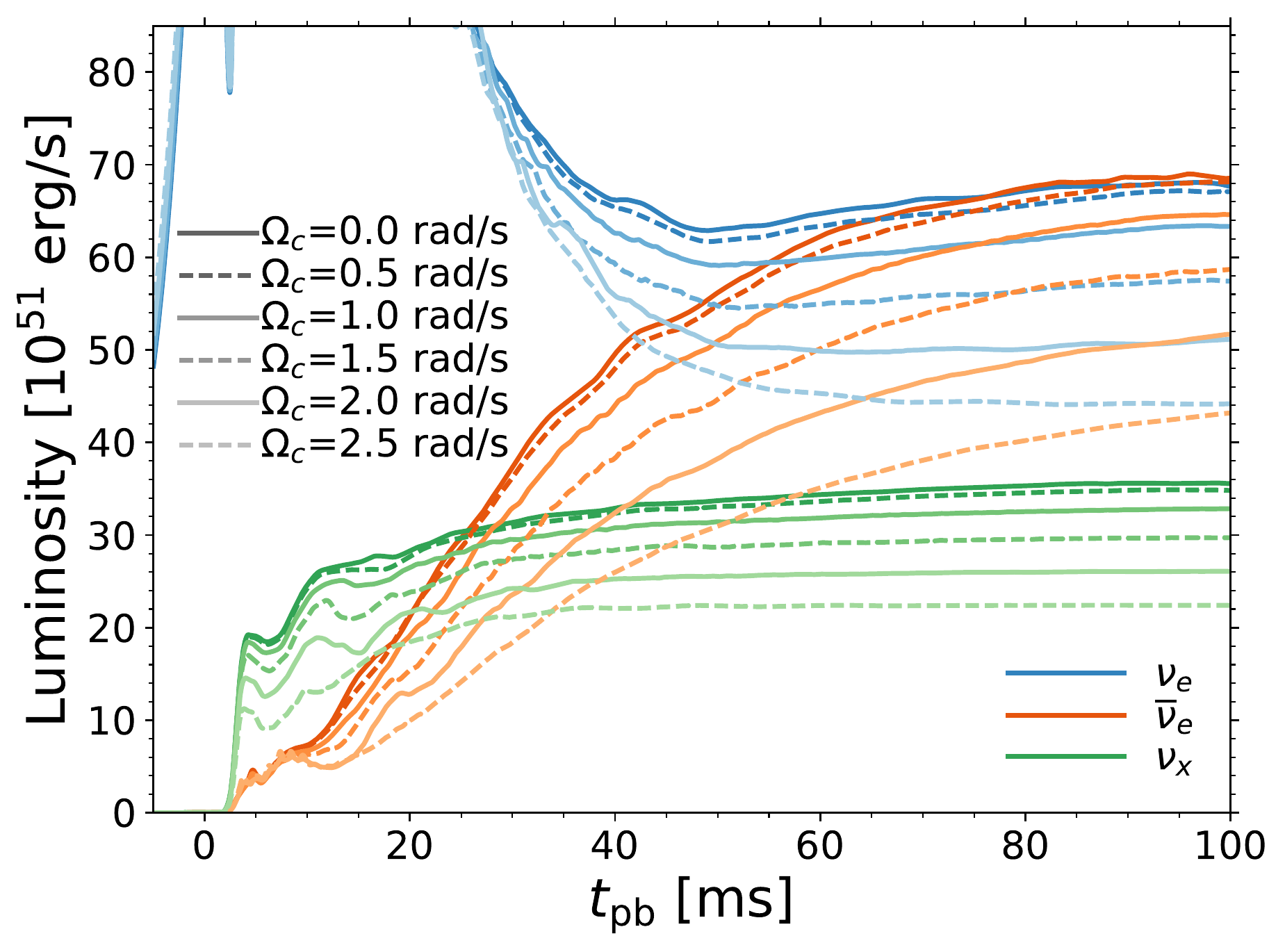}
\includegraphics[width=0.48\textwidth]{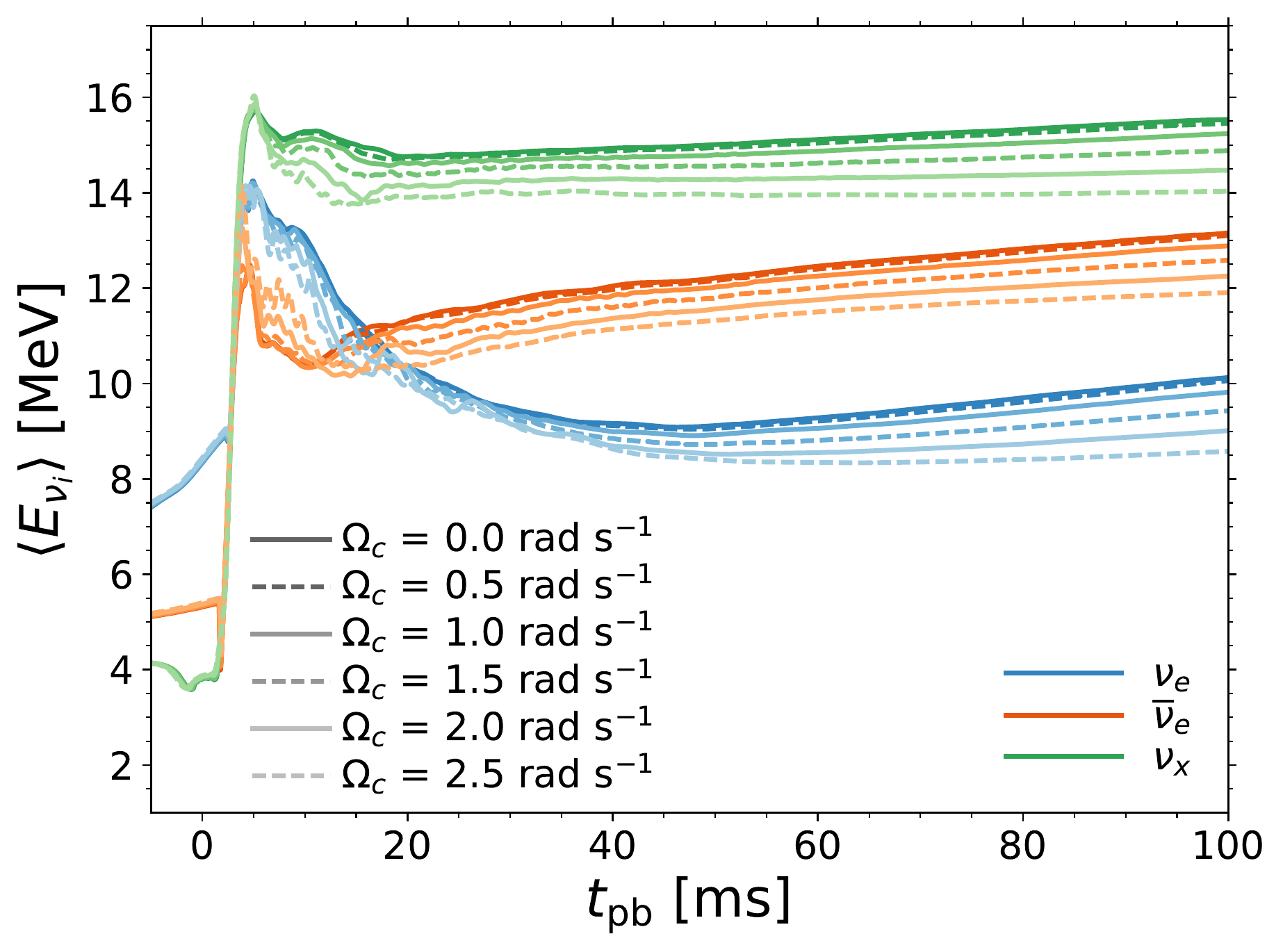}
\includegraphics[width=0.48\textwidth]{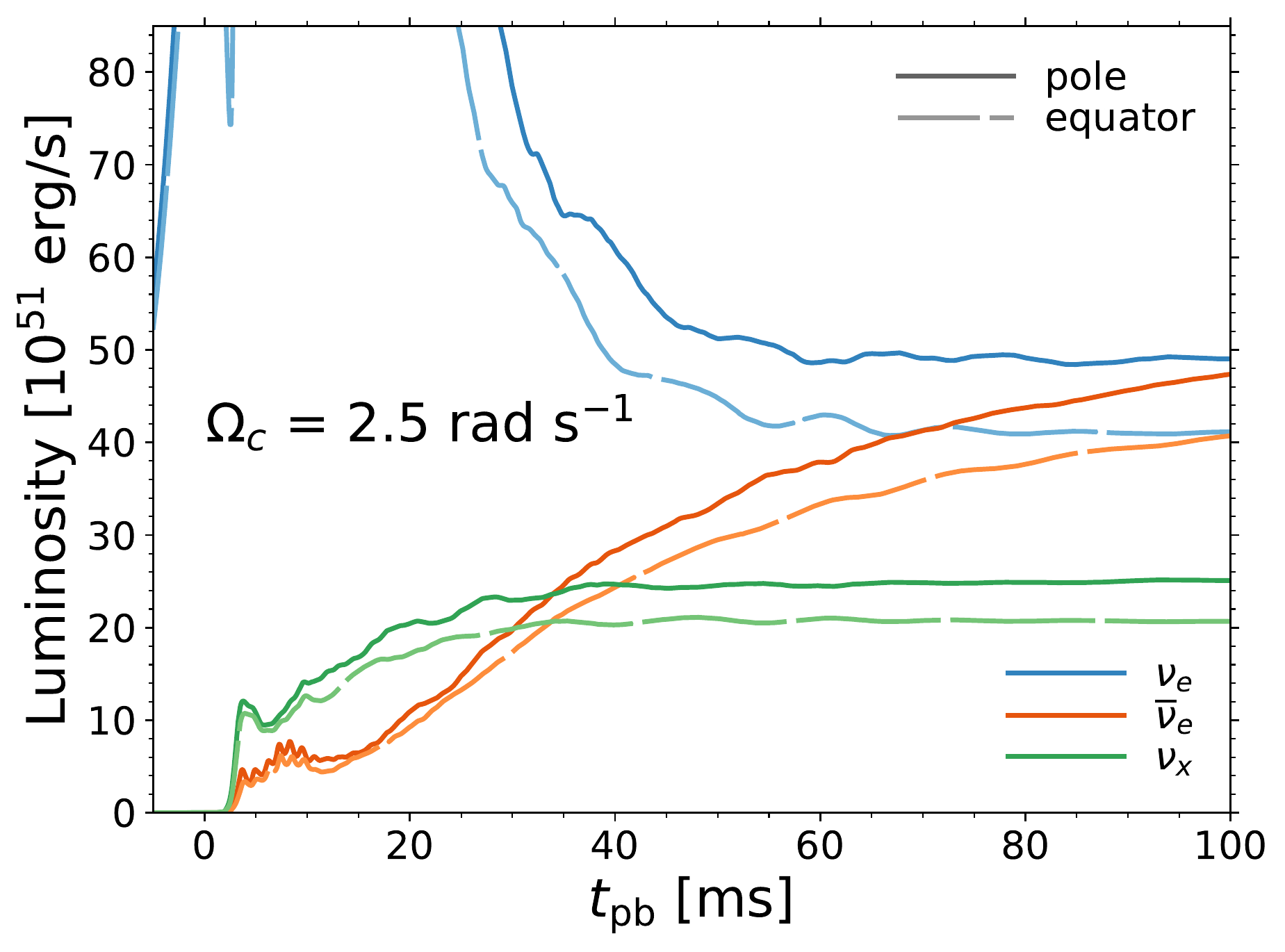}

\caption{Neutrino emission properties for various rotation rates and observing angles.  The top and middle panel shows the neutrino luminosity and average energy, respectively, for all rotations and for each neutrino species ($\nu_e$:\,blue, $\bar{\nu}_e$:\,orange, and $\nu_x$:\,green).  In the bottom panel we show the latitudinal dependence of the neutrino luminosity for $\Omega_c\!=\! 2.5$\,rad\,s$^{-1}$ by showing the luminosity seen by an observer both along the polar axis and on the equator. } \label{fig:allnulums}
\end{figure}

We perform a total of six simulations of rotating CCSNe in 2D axisymmetry using \texttt{FLASH} with a sequence of initial core rotation rates of $\Omega_c\!=\!\lbrace 0.0, 0.5, 1.0, 1.5, 2.0, 2.5\rbrace $\,rad\,s$^{-1}$. In this section, we present a brief overview of the simulations as a whole before exploring details of modes in particular simulations in the following section. 

The collapse times (from the start of the simulation to core bounce) range from 300\,ms for the non-rotating model to 326\,ms for the model rotating at $\Omega_c\!=\! 2.5$\,rad\,s$^{-1}$. We subtract off this time in all our results below. 

The rotation causes the collapsing core and subsequent PNS to become oblate and partially centrifugally supported. For example, at $\sim$40\,ms after bounce, the oblate iso-density contours at a density of $10^{12}$\,g\,cm$^{-3}$ have polar-to-equatorial radii ratios of $\sim$1, $\sim$0.98, $\sim$0.94, $\sim$0.88, $\sim$0.82, and $\sim$0.74 for our six simulations in order of increasing $\Omega_c$. The shock radii evolution is fairly similar in all models up to the end of the simulated time, $\sim$100\,ms, with only a mild rotational dependence. The mean shock radius at 100\,ms ranges from  155\,km in the non-rotating case to 165\,km in the fastest rotating case.  The small amount of turbulent motion that is present at this early time shows the expected (at least in 2D) dependence on rotation, that is an overall supression with higher rotation rates \cite{pajkos2019gw}.  The rotation itself slows down the accretion of matter onto the PNS, but this is a small effect in these simulations. The $\Omega_c=2.5$\,rad\,s$^{-1}$ simulation has a $\sim$3\% ($\sim$5\%) lower mass accretion rate, as measured at 500\,km, when compared to the non-rotating model at the time of bounce (at $\sim$100\,ms after bounce).

The added centrifugal support also reduces the gravitational binding energy released and consequently the emergent neutrino luminosity and neutrino average energy. In \fref{fig:allnulums}, we show the sky-averaged neutrino luminosity (top panel) and sky-averaged neutrino average energy (middle panel) for electron neutrinos (blue), antineutrinos (orange), and a characteristic heavy-lepton neutrino (green) for each of the rotation rates explored. The neutrino information was extracted at 500\,km. The electron neutrino neutronization burst is minimally impacted by the rotation. However, the remaining species have reduced emission for increasing rotation rates, as well as the electron neutrinos after the neutronization burst ($t \gtrsim 30$\,ms). The neutrino luminosity is reduced by at least 35\% for $\Omega_c\!=\! 2.5$\,rad\,s$^{-1}$ for all neutrino species at 100\,ms after bounce, while the corresponding neutrino average energy is reduced by at least $\sim$10\%. With increasing $\Omega_c$, the increase in the bounce time, and reductions in neutrino luminosity and neutrino average energy scale as $\Omega_c^2$ \cite{oconnor:11,fryer2000}.

The luminosities plotted in the top panel of \fref{fig:allnulums} are sky-averaged; however, there is also a latitudinal dependence of the neutrino luminosity, displayed in the lower panel of \fref{fig:allnulums}.  For illustration we only show the $\Omega_c\!=\! 2.5$\,rad\,s$^{-1}$ case, for which the neutrino emission has the strongest angular dependence. In the analysis that follows we consider both of these directions, pole and equator, which are constructed by averaging the emergent neutrino fields (to reduce numerical noise) at a radius of 500\,km from within 30$^\circ$ of the pole and $\pm$15$^\circ$ of the equator, respectively.

For the non-zero rotation rates, and especially evident for $\Omega_c\!=\! 1.0$\,rad\,s$^{-1}$ around 40\,ms after bounce and in the fastest rotating case ($\Omega_c\!=\! 2.5$\,rad\,s$^{-1}$) directly following bounce, we see small amplitude, high frequency oscillations imprinted on the neutrino luminosities and average energies. The luminosities and average energies are in phase, which has implications for the ability to detect these oscillations. For this work, we focus on the oscillations in the moderately rotating case, $\Omega_c\!=\! 1.0$\,rad\,s$^{-1}$. We note that the oscillations seen soon ($\sim\!5-10$\,ms) after bounce in the fastest rotating case are precisely the signal seen in \cite{ott2012correlated}.

\begin{figure}
\centering
\includegraphics[width=0.48\textwidth]{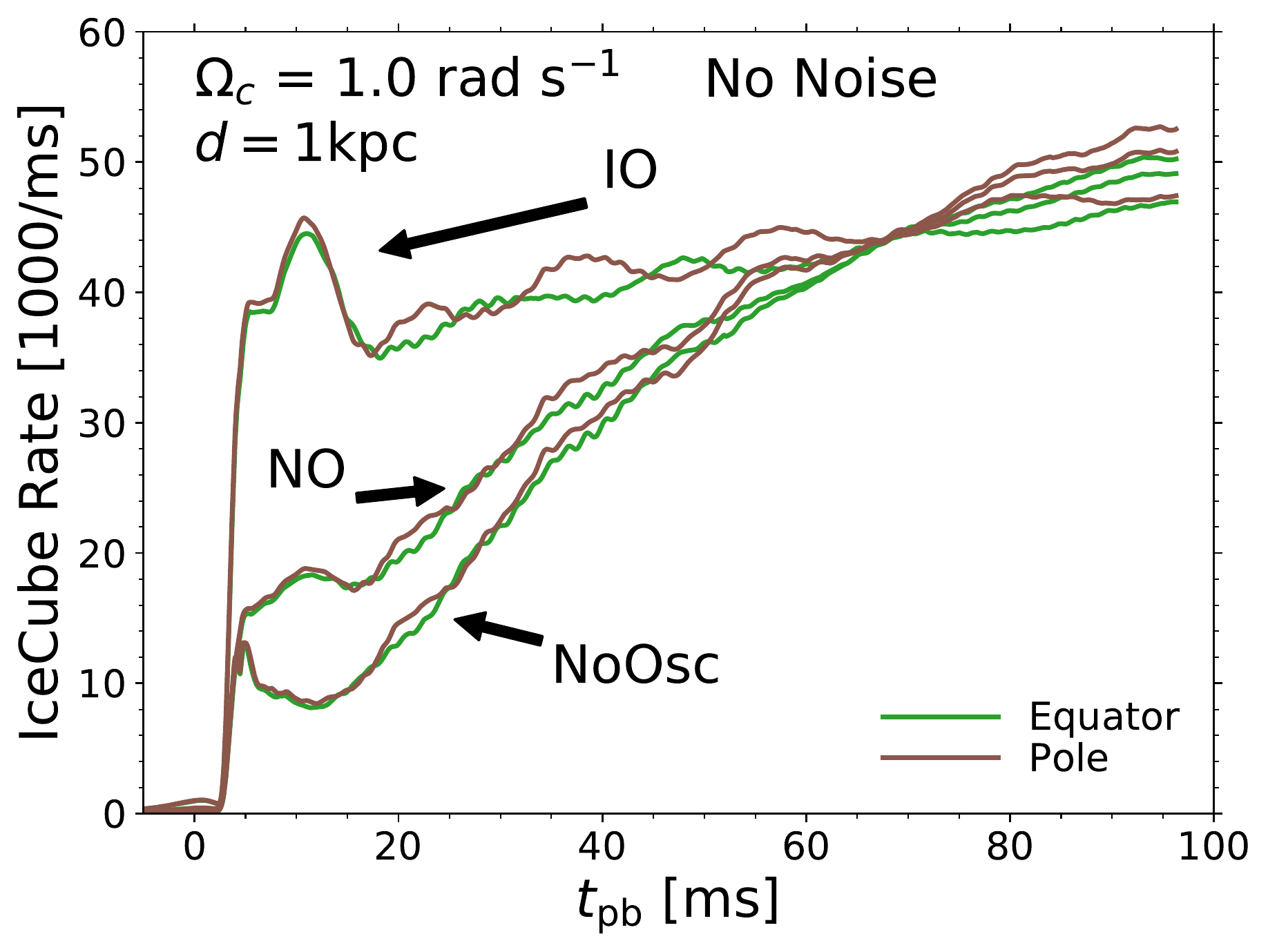}
\includegraphics[width=0.48\textwidth]{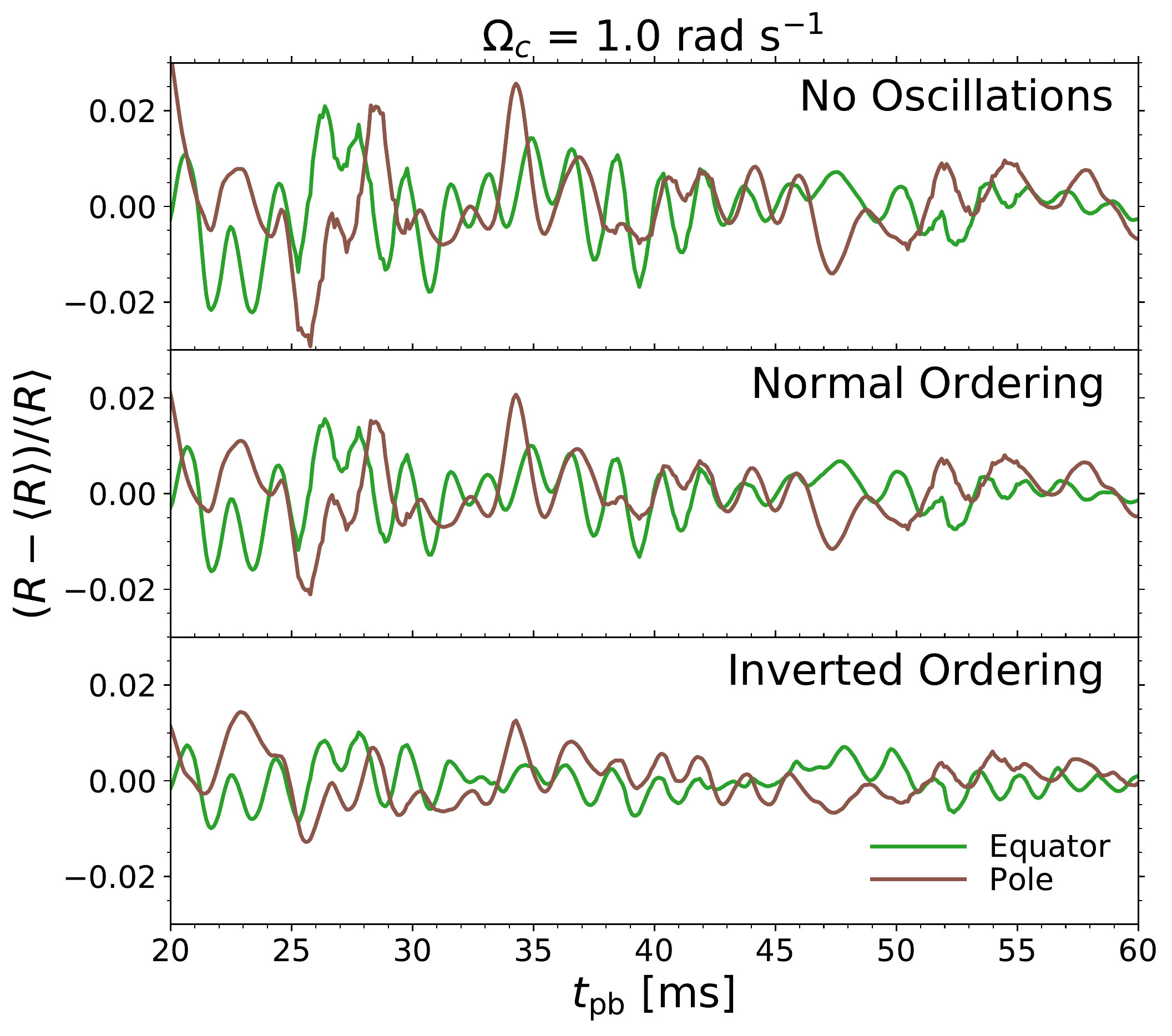}
\caption{Predicted event rates (top panel) in IceCube for various flavor oscillation scenarios (NoOsc: no oscillation effects, NO: normal ordering, IO: inverted ordering) and observer positions (green: equator, brown: pole) for the $\Omega_c\!=\! 1.0$\,rad s$^{-1}$ simulation located at 1\,kpc. In the bottom panel we show the high frequency content of the neutrino signal by showing the rates relative to a 5\,ms running average of the direction dependent signal (i.e. the top panel). The characteristic frequency matches the expectation from the GWs (see Fig.~\ref{fig:rot10_l2mode_Lnusky}), and the amplitude is 1-2\%.} \label{fig:rates}
\end{figure}

To infer the detectability of the neutrino signal (more details in \sref{sec:detect}), we use the SNOwGLoBES package \cite{scholberg:12}. SNOwGLoBES is a fast calculator for expected detection rates of CCSN neutrinos. Our reference neutrino detector is IceCube \cite{Achterberg:2006md,Abbasi:2011ss}, a cubic-kilometer-scale neutrino detector located at the geographic South Pole, which has been recently incorporated into the SNOwGLoBES code \cite{malmenbeckICRC}. The detection of CCSN neutrinos in IceCube comes primarily from the inverse beta decay reactions that arise from $\bar{\nu}_e$ interactions with the free protons in the ice.

For a galactic CCSN, IceCube will measure the rate evolution of the neutrino signal with the best statistical accuracy~\cite{Kopke:2017req}. In the top panel of \fref{fig:rates}, we show the SNOwGLoBES predicted rates in the IceCube neutrino telescope for the $\Omega_c\!=\! 1.0$\,rad\,s$^{-1}$ model located at a distance of 1\,kpc. In this figure we do not include IceCube dark rate noise or any statistical error due to counting statistics (although we do include these noise sources in the detectability analysis in \sref{sec:detect}).  We show the predicted rate for purely adiabatic MSW neutrino oscillations (ignoring any modification of the neutrino lightcurve due to neutrino-neutrino interactions) in both orderings [normal (NO)) and inverted (IO)] as well as assuming no oscillation effects (NoOsc). 
In the normal ordering, the $\bar{\nu}_e$ signal at Earth is a mixture of the original $\bar{\nu}_e$ signal ($\sim$70\%) and the original $\nu_x$ signal ($\sim$30\%).  For the inverted ordering, the $\bar{\nu}_e$ at Earth is almost completely the original $\nu_x$ ($\sim$100\%).
The green line shows the predictions based on an observer located on the equator while the brown is for an observer on the pole. 

To highlight the imprint of the oscillations, we show in the bottom panel of \fref{fig:rates} the neutrino rates relative to a 5\,ms running average of the corresponding rate from the top panel. Here we again separate the neutrino orderings and the equatorial and polar signals.  The typical high frequency content of the equatorial neutrino rates is $\sim 550$-575\,Hz (22-23 cycles over 40\,ms). As we show below, this is similar to the $h_+$ GW signal. However, the relative amplitude of the oscillations are only 1-2\% of the background neutrino signal.  We will require a close by CCSN in order to have enough statistics to observe this feature.

\begin{figure*}
\centering
\includegraphics[width=0.9\textwidth]{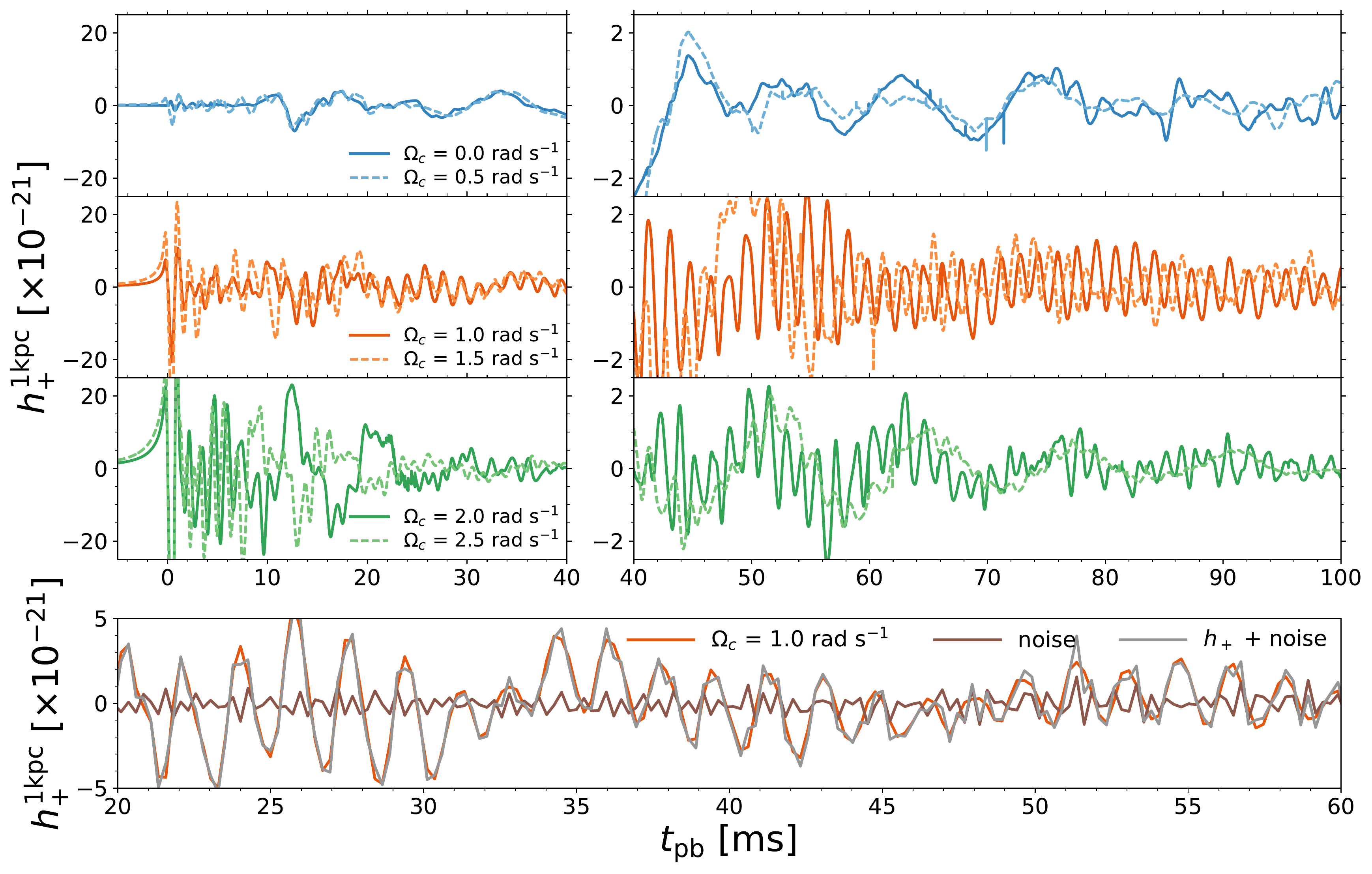}
\caption{Gravitational wave strains over the simulated time for the six progenitors explored, 0.0 and 0.5\,rad\,s$^{-1}$ (top panel, blue lines), 1.0 and 1.5\,rad\,s$^{-1}$ (middle panel, orange lines), and 2.0 and 2.5\,rad\,s$^{-1}$ (bottom panel, green lines). On the left, we show the first 40\,ms of GW data, while on the right we show an enlarged view for the remain 60\,ms.  In the lower plot, we show a subset of the $\Omega_c\!=\! 1.0$\,rad\,s$^{-1}$ data from 20\,ms to 60\,ms (orange), a realization of Advanced LIGO design sensitivity noise (brown), and the sum (grey).  All signals are scaled to 1\,kpc and are viewed on the
equatorial plane. Small glitches in the GW data near 50-70\,ms are due to the shock crossing mesh refinement boundaries.} \label{fig:gwwave_train}
\end{figure*}

The rapidly contracting, rotating, oblate spheroid also generates a GW signal.  As mentioned above, our simulations are axisymmetric and therefore the only non-zero GW signal is the $h_+$ polarization.  This peaks for an observer along the equator and vanishes for an observer at the pole. Notice that the neutrino luminosity is instead maximal at the pole and minimal at the equator, although with a weaker dependence. The characteristic GW signal of the rotating, collapsing core peaks at core bounce and subsequently rings down, over the course of $\sim$20\,ms as the core settles into its new equilibrium. Further on in the evolution the GW signal will become loud again when convection and turbulence kick in, although in rotating models this is expected to be muted relative to the non-rotating case \cite{pajkos2019gw}. We show the GW strain as a function of time for our six models in \fref{fig:gwwave_train} for an observer located on the equatorial plane.  With increasing rotation, this signal becomes more pronounced.  Persistent after this time are characteristic excited modes (which will be discussed in the following sections) which radiate GWs for the remainder of our simulations (up to 100\,ms after bounce).  Additionally, low frequency GWs are present even in the non-rotating simulations. For reference, the total energy radiated in gravitational waves (\cite{morozova2018gravitational}) up to 100\,ms after bounce is (in units of $M_\odot c^2$) $\sim8.5\times 10^{-10}$, $\sim1.1\times 10^{-9}$, $\sim3.5\times 10^{-9}$, $\sim1.0\times 10^{-8}$, $\sim3.9\times 10^{-8}$, and $\sim3.3\times 10^{-8}$ for our six simulations in order of increasing $\Omega_c$ from 0.0 rad\,s$^{-1}$ to 2.5 rad\,s$^{-1}$.

In the bottom panel of \fref{fig:gwwave_train}, we show a subset of the $\Omega_c\!=\! 1.0$\,rad\,s$^{-1}$ GW signal between 20 and 60\,ms after core bounce for a CCSN located at 1\,kpc (orange). Here we see a dominant and persistent frequency of $\sim$575\,Hz ($\sim$23 cycles over 40\,ms). This is consistent with the neutrino signal discussed above.  We also include in this panel a realization of Advanced LIGO design sensitivity noise (brown) \cite{pycbc}, and the resulting expected signal (grey).

%
%
\section{Mode Identification} \label{sec:modemeth}

In this section, we describe the procedure for identifying modes of oscillation of the system. The description we give here is very terse, and so we refer the interested reader to~\cite{westernacher2018turbulence}, where our analysis is described in great detail and exhaustively demonstrated.
The basic strategy is to compute the spectrum of linear modes of the system via perturbation theory, and then perform a matching between those modes and the full nonlinear simulation. The matching step is crucial, since the simulation tells us which modes are actually excited.

\subsection{Perturbative schemes} \label{sec:pertschemes}

We use a perturbative scheme in the relativistic Cowling approximation, as described in~\cite{torres2017towards}. The scheme assumes spherical symmetry and a coordinate system which accommodates our numerical setup of Euclidean spatial metric and vanishing shift vector. The lapse function is obtained from the effective relativistic gravitational potential $\Phi$ via $\alpha = e^\Phi$. 

The relativistic hydrodynamic equations are perturbed on top of the fixed background spacetime, and the solution ansatz uses spherical harmonics for the angular dependence, harmonic time dependence, and an unspecified radial profile which is solved for by outward radial integration of a system of ordinary differential equations in $r$. Spherically-averaged snapshots from our full nonlinear simulations are used as background solutions on which the perturbative calculations are performed. 

The radial displacement is prescribed to be a small number at the first off-origin grid point, and the transverse displacement is determined by a regularity condition in a neighborhood of the origin. This regularity condition was misreported in~\cite{torres2017towards,morozova2018gravitational} and subsequently corrected in~\cite{torres2018towards}. The error was pointed out in~\cite{westernacher2018turbulence} and found not to produce significant errors in computed mode frequencies. In~\cite{westernacher2018turbulence}, for simplicity of analysis the outer boundary condition was taken to be the vanishing of the radial displacement at $r$\,=100\,km.

In this work, we checked how the mode identification of~\cite{westernacher2018turbulence} changes when we instead use the outer boundary condition of~\cite{torres2017towards}, in which the radial displacement is taken to vanish at the position of the shockwave. The main difference is that the number of radial nodes increases, as anticipated in~\cite{westernacher2018turbulence}; the $l=2$, $m=0$, $n \gtrsim 2$ mode reported in~\cite{westernacher2018turbulence} reveals two additional nodes in the outer low-density region $r \gtrsim 90$ km. We explore various choices of boundary conditions in Appendix \sref{sec:modedurability}, and find broad robustness of our results.

\subsection{Mode function matching} \label{sec:modeextract}
Using the procedure described in \sref{sec:pertschemes}, we obtain the linear spectrum of the CCSN system at a given time. All of the information about the mode excitations is in the full nonlinear simulation itself. The determination of which modes are excited involves a best-fit matching procedure between the perturbative mode functions and the velocity data from the simulations. Perturbatively we solve for the displacement field, whereas we are comparing to velocity fields in the simulations; harmonic time dependence ensures that the two fields are proportional. In particular, the displacement field $\xi^i$ is related to the advective velocity perturbation $\delta v^{* i}$ and Eulerian velocity perturbation $\delta v^i$ via $\partial_t \xi^i = \delta v^{* i} = \alpha \delta v^i$, and harmonic time dependence means $\partial_t \xi^i \propto \xi^i$. Thus we compare $\xi^i/\alpha$ with the Eulerian velocity data in our simulations.

Prior to searching for the best-fit mode function, the velocity field of the star is processed through a time-varying spectral filter. This is more appropriate than a band-pass filter, since mode frequencies can change in time. The spectral filters are chosen to extract motions identifiable in the velocity field itself, rather than the GW signal, since not all modes will generate significant GWs (but may do so in rotating stars, where the modes acquire quadrupolar deformations). The spectral filter mask is a time-varying top-hat window, drawn manually on the spectrograms of the velocity field based on visual identification of excited features. In the future it would be desirable to automate this process, to increase reproducibility. But such automation may require something akin to machine learning, which is well beyond our scope. Our filter kernel masks are displayed on a sampling of velocity spectrograms in Appendix \sref{sec:kernels}. In~\cite{westernacher2018turbulence} the analysis was found not to be sensitive to shrinking the kernel masks in their frequency extent by a factor of 2.

The spectrally-filtered velocity field is then decomposed in a vector spherical harmonic basis. In this way, we obtain a set of simulation velocity fields we denote schematically as $^{l}\vec{v}_{\sigma,\mathrm{sim}}$, where $l$ is the spherical harmonic number and $\sigma$ is the (average) frequency of the spectral filter used. 

To compare with perturbative mode functions, which we denote as $^{l} \vec{v}_{\sigma^\prime,\mathrm{pert}}$ (where $\sigma^\prime$ is its frequency), we use a measure of difference defined as
\begin{eqnarray}
\Delta \equiv \sqrt{\sum \left( {}^l \vec{v}_{\sigma^\prime,\mathrm{pert}} - {}^l \vec{v}_{\sigma,\mathrm{sim}} \right)^2 }, \label{eq:Delta}
\end{eqnarray}
where the sum is over radial points on the numerical grid. Mathematically this is a Frobenius norm. The best-fit mode function for $^{l}\vec{v}_{\sigma,\mathrm{sim}}$ is found by minimizing $\Delta$ over the perturbative mode spectrum, which is parameterized by a discrete set of frequencies $\sigma$ whose mode functions satisfy the outer boundary condition. Prior to matching, both velocity fields are normalized by their $L_2$-norms, since we only wish to compare their shapes.

After identifying the active modes in the non-rotating case, we can repeat the extraction of $^{l}\vec{v}_{\sigma,\mathrm{sim}}$ in the rotating cases. We do not compare with perturbative mode functions in rotating models, since our perturbative scheme is not valid there. We instead observe the progressive change in the mode eigenfunctions across the different rotation cases~\cite{friedman2013rotating,dimmelmeier2006non}. We use a similar measure of difference as $\Delta$, except between the simulated velocity fields between two adjacent models in the rotating sequence, i.e.~between ${}^l\vec{v}_{\sigma^\prime,\mathrm{sim 1}}$ and ${}^l\vec{v}_{\sigma,\mathrm{sim 2}}$ where simulations 1 and 2 are adjacent in the rotating sequence. This amounts to following the continuity of mode functions along the sequence of rotating models.

In practice, we apply a mass density weighting $\sqrt{\rho}$ to the velocity fields and mode functions prior to matching. This acts to discount fluctuations in the simulated data occurring further from the core which are not representative of a linear mode of the system. In this work we display mode functions with a weaker $\rho^{1/4}$ weighting, which is less forgiving but allows for a visual inspection of the mode functions at larger radii, so long as we also smooth the simulated data.

In rotating models, mode functions are no longer pure spherical harmonics. In~\cite{westernacher2018turbulence}, we identified deformations of modes via consistency with parity selection rules, unified exponential decay rates and oscillation frequencies, and expectations from first and second order perturbation theory in rotation~\cite{friedman2013rotating}. The identification of mode deformations is not a crucial part of this work, so we leave the details of the procedure to~\cite{westernacher2018turbulence}. A powerful method in~\cite{dimmelmeier2006non} called \emph{mode recycling} was used to converge toward the mode function of rotating stars, but is not available in our context. One simulates the star with an initial perturbation corresponding to an educated guess for the mode function of interest, which due to its inaccuracy will excite several unwanted modes. By applying spectral filters to the velocity field in the star, a more accurate trial mode function for the target mode can be extracted and used as an initial perturbation in a second simulation. This process is repeated until the initial perturbation results in a clean excitation of the target mode, with unwanted modes highly suppressed. We cannot use mode recycling since our target modes are excited by core bounce, which we do not attempt to manipulate.
%
%

\section{The Multimessenger Imprint of a Proto-Neutron Star Mode} \label{sec:dualimprint}

%
%

\begin{figure*}[htbp]
\centering
\includegraphics[width=0.9\textwidth]{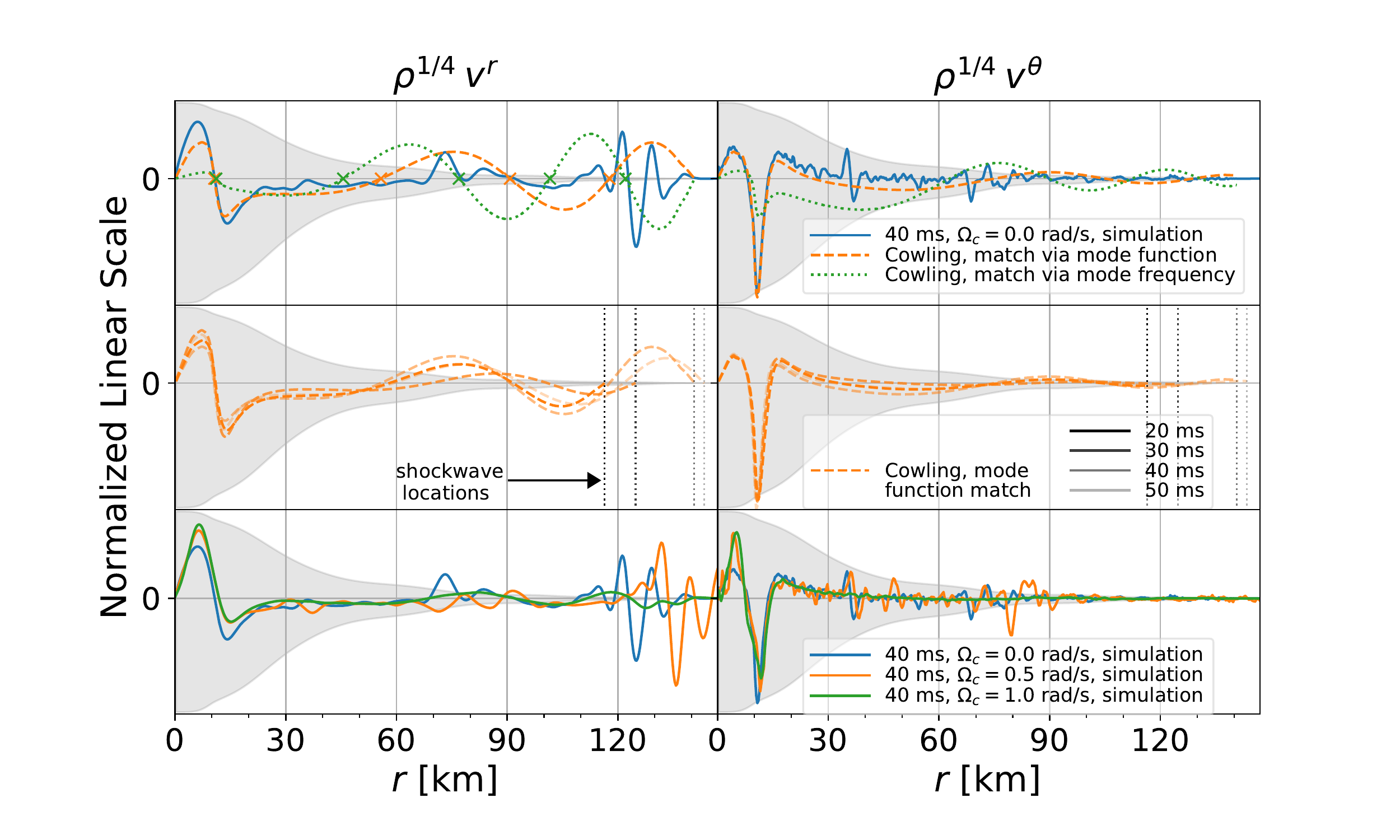}
\caption{Simulation data and best-fit $l$\,=\,2 modes in the Cowling approximation. Radial and angular components are displayed in the left and right columns, respectively. The radial simulation data has been smoothed with a Gaussian of width $1.5$ km, in order to allow a visual inspection of the peaks and troughs across $\sim 50$\,-$110$\,km. The angular component has not been smoothed. The shaded region displays the fraction of energy external to radius $r$, which was computed using the Cowling perturbative mode function at $40$ ms and the simulated, spherically-averaged density $\rho$. The shaded region is intended to indicate where the quality of mode match is most important. Both the simulation data and the perturbative modes have been normalized by their $L_2$-norms, just as they are during the mode function matching procedure. Top Row: Snapshots of the $\rho^{1/4}$-weighted velocity field from the simulation in the neighborhood of $40$ ms, as well as the perturbative mode functions in the Cowling approximation which are matched via best-fit mode function (dashed lines) and via best-fit mode frequency (dotted lines). Radial nodes of the perturbative mode functions are indicated with crosses. The additional zero-crossings in the simulated data we interpret as noise. Middle Row: Perturbative mode functions only, at varying times. Shockwave locations are indicated with vertical dotted lines. The number of nodes $n$ over the whole domain increases from $3$ to $4$ as the outer shockwave expands. Bottom Row: Simulation snapshots in the neighborhood of $40$ ms taken from the $\Omega_c\! =\! \left\lbrace 0.0,0.5,1.0 \right\rbrace$ rad\,s$^{-1}$ models. The band masks from which these snapshots are taken are those which yield maximal mode function continuity across the model sequence.} \label{fig:500Hz_modes}
\end{figure*}
In this section we implicate an $l\!=\!2$, $m\!=\!0$, $n\!\gtrsim\!2$ mode in producing prominent frequency peaks in both GWs and neutrinos for the $\Omega_c\!=\! 1.0$\,rad\,s$^{-1}$  model. We write $n \gtrsim 2$ despite the fact that $n\!=$\,3-4 if the nodes are counted all the way out to the shock wave (depending on the exact time), because the innermost 2 nodes exist more clearly within the PNS proper, whereas the outermost 1-2 nodes are in a low-density region. It is important to be explicit about this, since~\cite{sotani2016gravitational,morozova2018gravitational} placed the outer boundary condition at the PNS surface, whereas~\cite{torres2017towards} placed it at the shock wave. We find robustness of our mode identification on various boundary conditions in \sref{sec:modedurability}. In~\cite{sotani2019dependence}, results assuming both boundary conditions were compared. Thus, comparing the node counts with those works requires distinguishing between the nodes interior and exterior to the PNS.

In \fref{fig:500Hz_modes} we display the mode function matching. The shaded regions indicate the total mode energy exterior to $r$, and is intended to convey that the mode function matching is most significant in the inner $\sim 30$ km. This energy is obtained by integrating $\rho[\eta_r^2 + l(l+1)\eta_\theta^2/r^2]$ (see~\cite{torres2017towards}) from $r$ to the outer boundary, and then normalizing to 1. The top row of \fref{fig:500Hz_modes}  compares the simulation data with the best-fit perturbative modes for the $\Omega_c\!=\! 0.0$\,rad\,s$^{-1}$ model. Two best-fit modes are displayed, one fit according to mode function (dashed lines, frequency $423$ Hz), and the other fit according to mode frequency (dotted lines, frequency $515$ Hz). The mode rings at $\sim 490$ Hz in the simulation. All plots are normalized by their $L_2$-norms. Matching according to mode frequency yields a mode function which poorly represents the excitation observed in the simulation. Instead, matching via mode function yields a much better representation of the simulation, and allows for a convincing identification of radial nodes $n$, which would be very challenging with the simulation data alone. Nodes of the perturbative mode functions are indicated with crosses. 

The middle row of \fref{fig:500Hz_modes} compares best-fit perturbative mode functions at different times, which vary largely due to the movement of the location of the outer boundary condition (shockwave) during that time. The total number of radial nodes is seen to increase as the shockwave moves outward. 

The bottom row of \fref{fig:500Hz_modes}  compares simulation data across the $\Omega_c\!=\! \lbrace 0.0,0.5,1.0 \rbrace$ rad\,s$^{-1}$ models for the best-fit band masks. This illustrates our following of the mode via continuity of its mode function. Strikingly, the radial component in the $\Omega_c\!=\! 1.0$\,rad\,s$^{-1}$ model has clear zero-crossing behavior that is well-captured by the perturbative mode function of the non-rotating model, which suggests that the radial nodes have not shifted significantly. The clearer zero-crossing behavior we attribute to the larger excitation of the mode.

One usually regards as $p$-modes those modes which occur to the right of the minimum of the $n(f)$ curve (i.e., the radial node count as a function of mode frequency). That branch of modes has increasing frequency with increasing $n$. To the left of the minimum of $n(f)$ are $g$-modes, which have decreasing frequency with increasing $n$ (see eg.~\cite{torres2017towards,torres2018towards}). We are unable to determine whether the best-fit perturbative mode function in \fref{fig:500Hz_modes} in the non-rotating model is a $p$-mode or $g$-mode, since the best-fit mode occurs too close to the minimum of the $n(f)$ curve (see eg. Fig. 17.2 in~\cite{westernacher2018turbulence}). However, modes with smaller $n$ do not appear to exist at the times analyzed.

\begin{figure}[htbp]
\centering
\hspace{0cm}\includegraphics[width=0.4\textwidth]{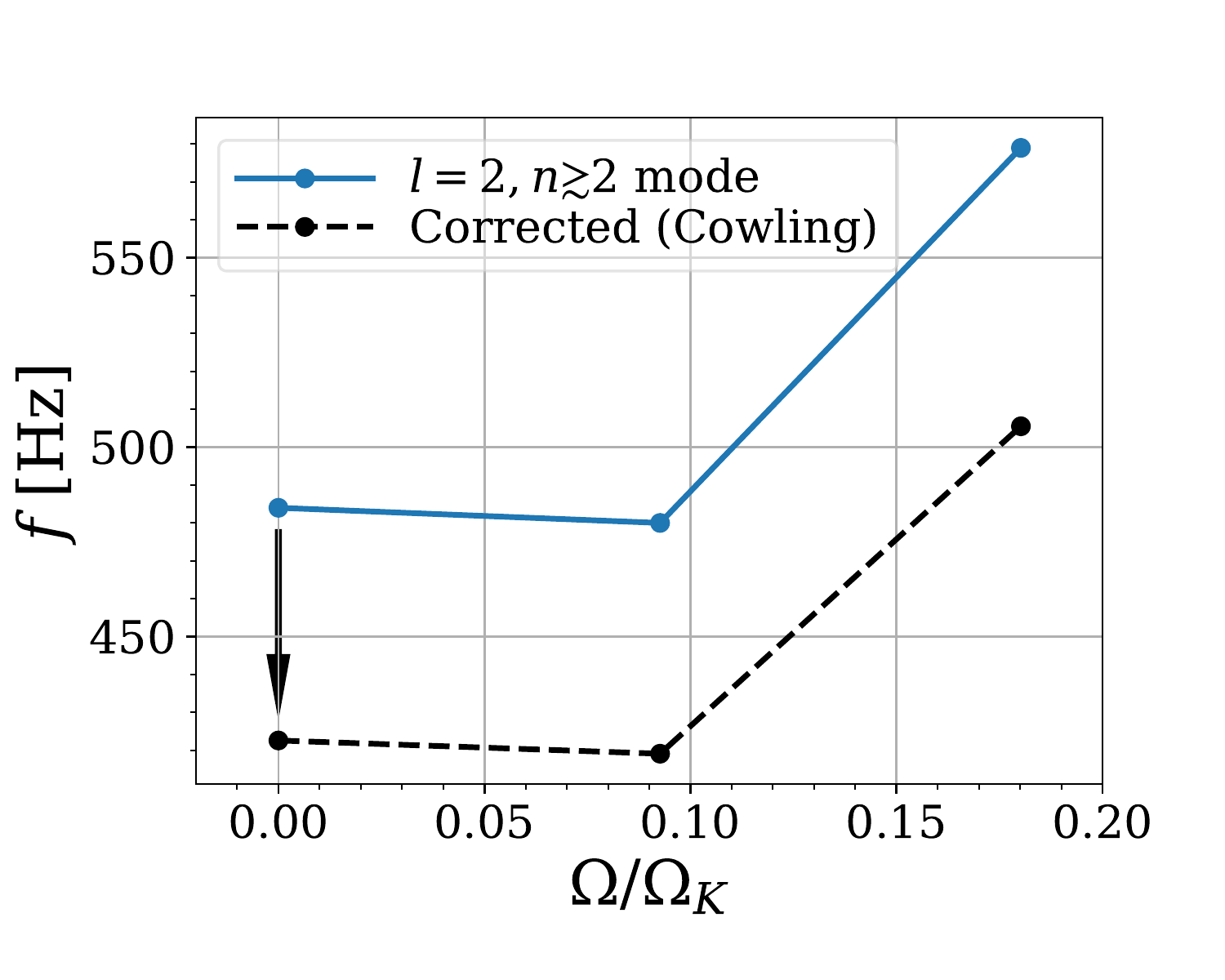}
\caption{The frequency of the $l\! =$\,2, $n\!\gtrsim$\,2 mode across the entire sequence of rotating models. $\Omega/\Omega_K$ is computed at $40$\,ms and averaged over the innermost $30$\,km, where $\Omega_K$ is the Keplerian frequency at $30$\,km. The frequencies (blue) are extracted from the $(l\! =$\,2$:\hat{r})$ component in each model. Corrected frequencies (black) are also shown, where we have scaled them down by $\sim 15\%$ of in order to match the frequency of the best-fit mode function obtained in the Cowling approximation for the non-rotating model.} \label{fig:l2mode_f_vs_Omega}
\end{figure}

In the $\Omega_c\!=\! 1.0$\,rad\,s$^{-1}$ model, the $l\!=$\,2, $n\!\gtrsim$\,2 mode picks up $l\!=$\,1,3 deformations with consistent parity, as well as with amplitudes consistent with expected leading order effects in rotation. The mode's frequency in the simulation, measured as an average over the mode's band mask, rises modestly from $\sim 490$ Hz to $\sim 570$ Hz in the $\Omega_c = 1.0$ rad\,s$^{-1}$ case. We display the change in mode frequency in \fref{fig:l2mode_f_vs_Omega}, together with a downward correction to the Cowling value in the non-rotating model\footnote{Note we use the correction factor coming from the mismatch in frequency between the non-rotating simulation and its best-fit mode function. The correction factor may vary as a function of rotation. However, since the mode ringing in the $\Omega_c\!=\! 1.0$\,rad\,s$^{-1}$ case is at a modestly different frequency than in the non-rotating case, we expect the correction factor is similar there.}. Since the central density of the system is decreasing as rotation increases, one would instead expect the frequency of this mode to decrease if the frequency scaled as $\sqrt{G \rho}$. It is unclear whether the increase in frequency is a result of the approximations being employed in the simulation, as explored in Appendix \sref{sec:FLASHTOV_test}, or whether the expected scaling $\sim\!\sqrt{G \rho}$ does not apply. 

The best-fit mode function in \fref{fig:500Hz_modes} has a frequency of $\sim\! 420\,$Hz, roughly $15\%$ lower than the frequency observed in the simulation itself. The lower frequency comes from a calculation in the Cowling approximation, which has been observed almost always to overestimate the true frequency of modes (i.e. the frequency when using full general relativity), see eg.~\cite{font2002three,cerda2008new,zink2010frequency, chirenti2015fundamental,mendes2018new}.\footnote{However, see the fundamental radial mode appearing in Fig.~11 of~\cite{torres2018towards} for an apparently glaring exception.} One expects this kind of systematic bias whenever a mode results in density fluctuations, since overdensities would backreact on the spacetime to produce an attractive influence in full general relativity (GR), thereby slowing the return to equilibrium. In the Cowling approximation, this backreaction is neglected. Thus, we expect that $420\,$Hz is as an upper bound on the true frequency of the mode on the CCSN background produced in our \texttt{FLASH} simulations. Based on the tests in Appendix \sref{sec:FLASHTOV_test}, one may expect the true frequency to be of order several $\%$ lower than this upper bound.

In order to narrow down the cause of the overestimated mode frequency in the simulations, we also perform a TOV oscillation test in the \texttt{FLASH} implementation in Appendix \sref{sec:FLASHTOV_test}. In this test, all of the physics has been eliminated except hydrodynamics and gravity. The TOV star is more compact than the PNS, and thus is a more demanding system. We find directionally consistent results, namely that the TOV mode frequencies are overestimated with respect to the Cowling values (except for the fundamental radial mode, which we have not focused on). This test implicates the lack of a GR metric in the hydrodynamics as a cause of the frequency overestimation. In particular, the solver lacks a lapse function in the hydrodynamic fluxes, and uses the pseudo-Newtonian gravitational potential~\cite{Keil1997,rampp2002radiation,marek2006exploring,couch13,o2018two}. The absence of densitization of the fluid variables by the metric determinant may also play a role. 

\begin{figure*}
\centering
\hbox{
\hspace{-1.0cm}\includegraphics[width=1.1\textwidth]{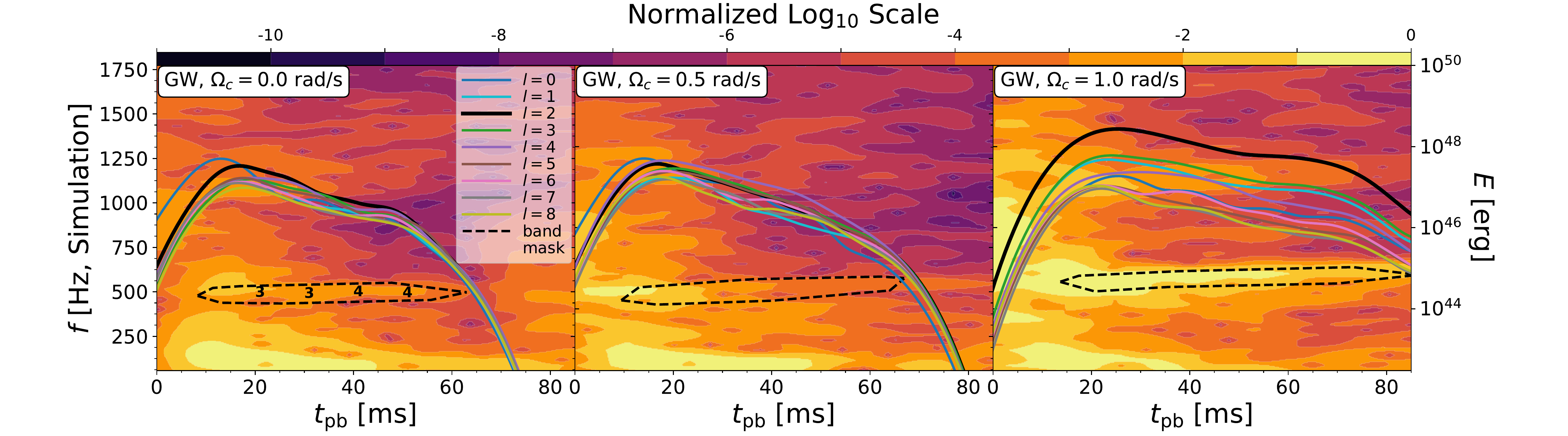}
}
\caption{Angular decompositions of the energy of the $\sim 500$ Hz band mask of the velocity field in the star (solid lines), for the $\Omega_c\!=\!\left\lbrace 0.0,0.5,1.0 \right\rbrace$\,rad\,s$^{-1}$ models (left to right). The $l\!=$\,2 component of the kinetic energy in the mask is plotted with a bold line. The contour plots display the GW spectrograms (with power spectrum scaling) on a normalized logarithmic scale, plotted against the simulation frequency axis; the true frequency of the mode contained in the mask is closer to $\sim 420$ Hz. The band masks are displayed as well (dashed lines). The energies have been smoothed with a Gaussian of width $3$ ms. The node counts of the best-fit mode function are displayed at their respective times for the $\Omega_c\!=\! 0.0$\,rad\,s$^{-1}$ case (left panel). In the $\Omega_c\!=\! 1.0$\,rad\,s$^{-1}$ case, the $l\!=$\,1 and $l\!=$\,3 components of the energy have distinguished themselves, and it was argued in~\cite{westernacher2018turbulence} that they are deformations of the $l\!=$\,2 mode occurring at first order in rotation.} \label{fig:500Hz_bands}
\end{figure*}

Writing $\alpha = e^\Phi$, we can estimate the lapse at the center of the star during the ringdown as $\sim \, 0.8$. Since this is the smallest value of the lapse in the system, this value provides an estimate of the maximum effect of the absence of the lapse on the mode frequency. This maximum effect of $20\%$ is consistent with the observed mismatch of $\sim 15\%$. However, the variability in the degree of overestimation in Appendix \sref{sec:FLASHTOV_test} suggests that the lapse is not the sole cause. Indeed, the fact that the only mode whose frequency \emph{improves} with respect to the Cowling value is the fundamental radial mode implicates the effective GR potential as well, since it was designed in spherical symmetry and so one would expect an improvement of the most dominant radial dynamics. The comparison in~\cite{muller2013new} between full GR and the effective GR potential focused on longer time scales of $\sim$ seconds, and they also observe overestimated frequencies\footnote{However, the comparisons in~\cite{muller2013new} do not involve comparisons of the mode functions, and therefore one does not actually know whether the same mode is being compared between the simulations using full GR and the effective potential.}. The TOV migration test has also been observed to produce stellar oscillations at about double the frequency as that observed in full GR~\cite{marek2006exploring,o2018two}.

One interesting possibility is that the mode excitation is moreso dependent on frequency rather than mode function. For example, if the mode excitation mechanism has a characteristic driving frequency, then it will tend to excite modes with resonant frequencies. In this case, in a full GR simulation one would still observe excitation of modes at similar frequencies as in a pseudo-Newtonian simulation, but the actual modes that are excited would be different. All of these observations emphasize the importance of using a mode function matching procedure rather than mode frequency matching, and doing so in a comparison between full GR and pseudo-Newtonian approaches.

The mode was followed to the $\Omega_c\!=\! 2.5$\,rad\,s$^{-1}$ model in~\cite{westernacher2018turbulence}. Upon reanalysis, and via the inclusion of an $\Omega_c\!=\! 1.5$\,rad\,s$^{-1}$ case, we make a more conservative conclusion in this work. Namely, we find the best-fit frequency bands going from $\Omega_c$=\,1.0$\rightarrow$1.5$\rightarrow$2.0\,rad\,s$^{-1}$ imply a rapid non-monotonic change in frequency, which is not expected on the basis of first or second order rotational effects, and therefore indicates that we are losing track of the mode. This may be partly due to the fact that the velocity field in the mode's expected band mask for $\Omega_c > 1.0$ rad\,s$^{-1}$ exhibits significant deviations from harmonic time dependence. The velocity field instead acquires a mixed character of harmonic and traveling-wave time dependence, and therefore becomes difficult to follow to higher rotation without a more sophisticated analysis strategy. This is one of the difficulties of not having fine control over the perturbations applied to the star, as one has for example when using full nonlinear simulations to study linear modes of rotating stars by carefully designing the applied perturbations~\cite{friedman2013rotating}.

Our focus here is instead on the $\Omega_c\!=\! 1.0$\,rad\,s$^{-1}$ model, where the mode has been followed well and the neutrino emission properties show a clear imprint from the mode. We emphasize that the analysis involved in following the mode across models is a separate methodology from the mode identification via mode function matching in the non-rotating model.

In \fref{fig:500Hz_bands}, overlaid on the GW strain spectrograms, we display an angular decomposition of the time-varying energy of the simulated velocity fields in the band masks of interest. The node counts of the best-fit perturbative mode function at $40$ ms are indicated for the non-rotating model. The $l\!\!=$\,2 component becomes highly distinguished in the $\Omega_c\!=\! 1.0$\,rad\,s$^{-1}$ model, as well as the $l\!=$\,1,3 deformations of the mode occurring at first order in rotation.

\begin{figure*}
\centering
\hbox{\hspace{-1cm}\includegraphics[width=1.1\textwidth]{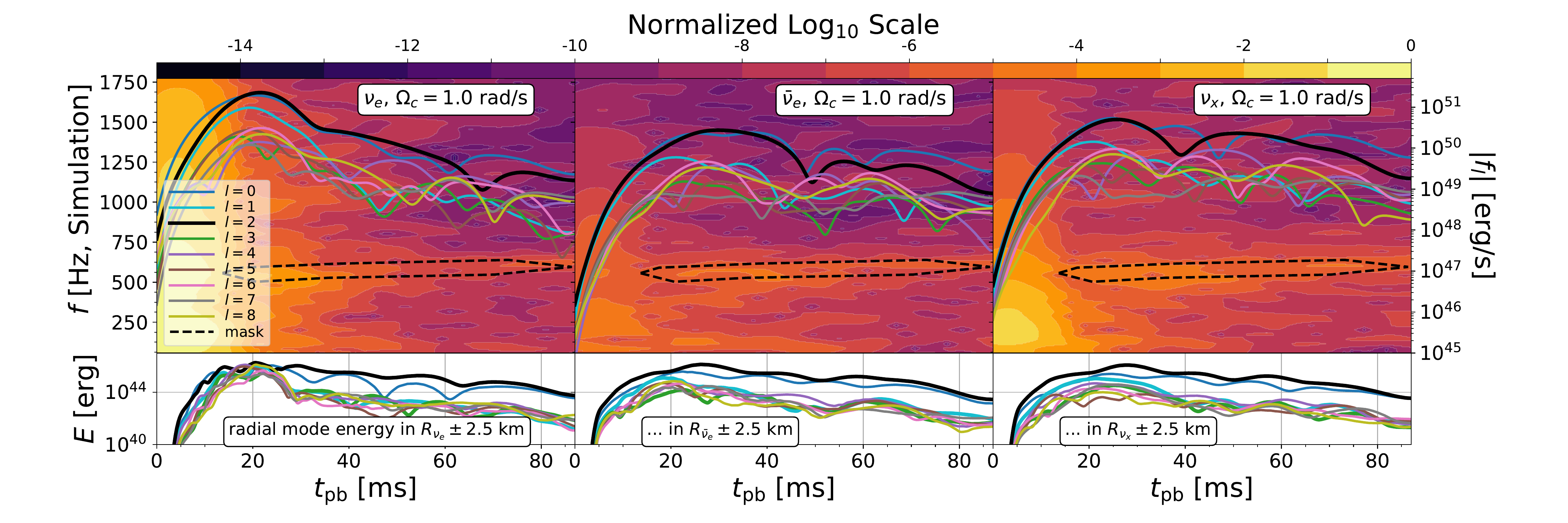}}
\caption{Top Row: Spherical harmonic decompositions of the neutrino luminosities on a sphere at $500$ km for species $\nu_e$ (left), $\bar{\nu}_e$ (centre), $\nu_x$ (right) plotted on top of spectrograms of the oscillating part of the sky-averaged neutrino light curves. The band mask is displayed, and is coincident with a prominent emission feature in the spectrograms. The vertical frequency axis is according to the simulation, which we argue overestimates the true frequency of the mode in the band mask by a factor $\sim 0.87^{-1}$. The neutrino light curves along each direction on the sky have had a Gaussian smoothing subtracted, and underwent the same spectrogram filtering as we applied to the velocity field using the band mask shown. The resulting time series were then decomposed angularly to obtain spherical harmonic coefficients $f_l$ at each time. The absolute value $|f_l|$ is then smoothed with a Gaussian of width $10$ ms before plotting. Bottom Row: The corresponding radial energy of the PNS in the band mask, integrated over a $5$ km width radial shell centered on the respective neutrinospheres. The harmonics $l\!=$\,0,2 stand out in all cases.} \label{fig:rot10_l2mode_Lnusky}
\end{figure*}

In \fref{fig:rot10_l2mode_Lnusky} we demonstrate for the $\Omega_c\!=\! 1.0$\,rad\,s$^{-1}$ model that the emission pattern of neutrinos on the sky at the frequencies inside the band mask is coincident with the angular distribution of radial kinetic energy in the star within a $5$ km width shell around the neutrinospheres. The top row contour maps show the neutrino luminosity spectrograms, where a moving average has been subtracted first in order to accentuate the oscillations. Overlaid on the contour maps are the coefficients $|f_l|$ of the angular decomposition of the spectrally-filtered neutrino emission on the sky, as a function of time. The bottom row plots the radial kinetic energy of the star near the neutrinospheres, also angularly decomposed. We observe that the $l\!=$\,0,2 components of the emission pattern on the sky are dominant, and those two components are also distinguished in the kinetic energy around the neutrinospheres. This is evidence that the mechanism of imprint of the mode onto the neutrino luminosity is via periodic variations of the neutrinospheres by the mode. The modulations in the neutrino luminosity therefore carry asteroseismological information regarding the mode amplitude in the vicinity of the neutrinospheres, which is complimentary to the deeper information carried out by GWs.

%
%
\section{Multimessenger Detectability} \label{sec:detect}

 \begin{figure*}
\centering
\includegraphics[width=0.306\textwidth]{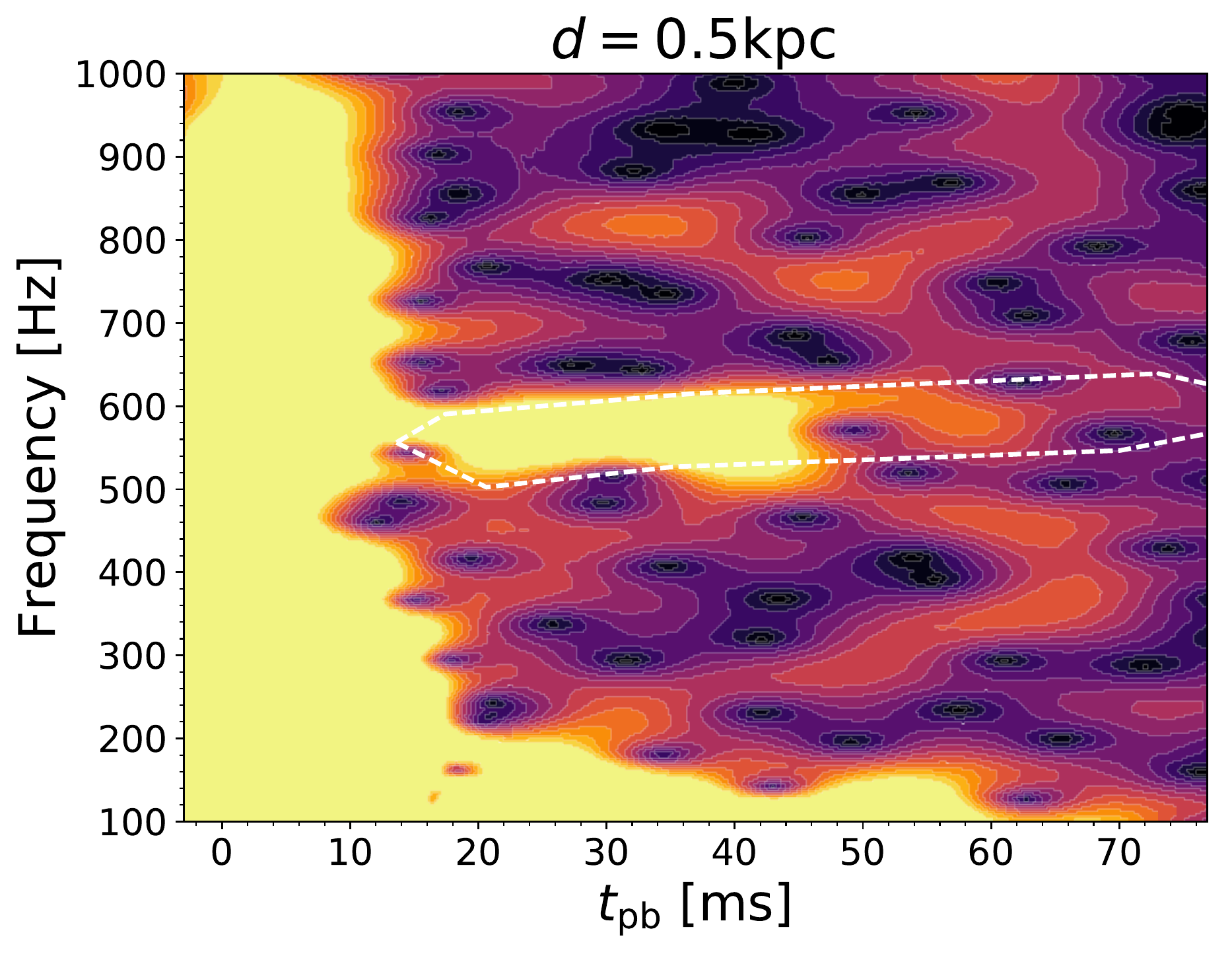}
\includegraphics[width=0.306\textwidth]{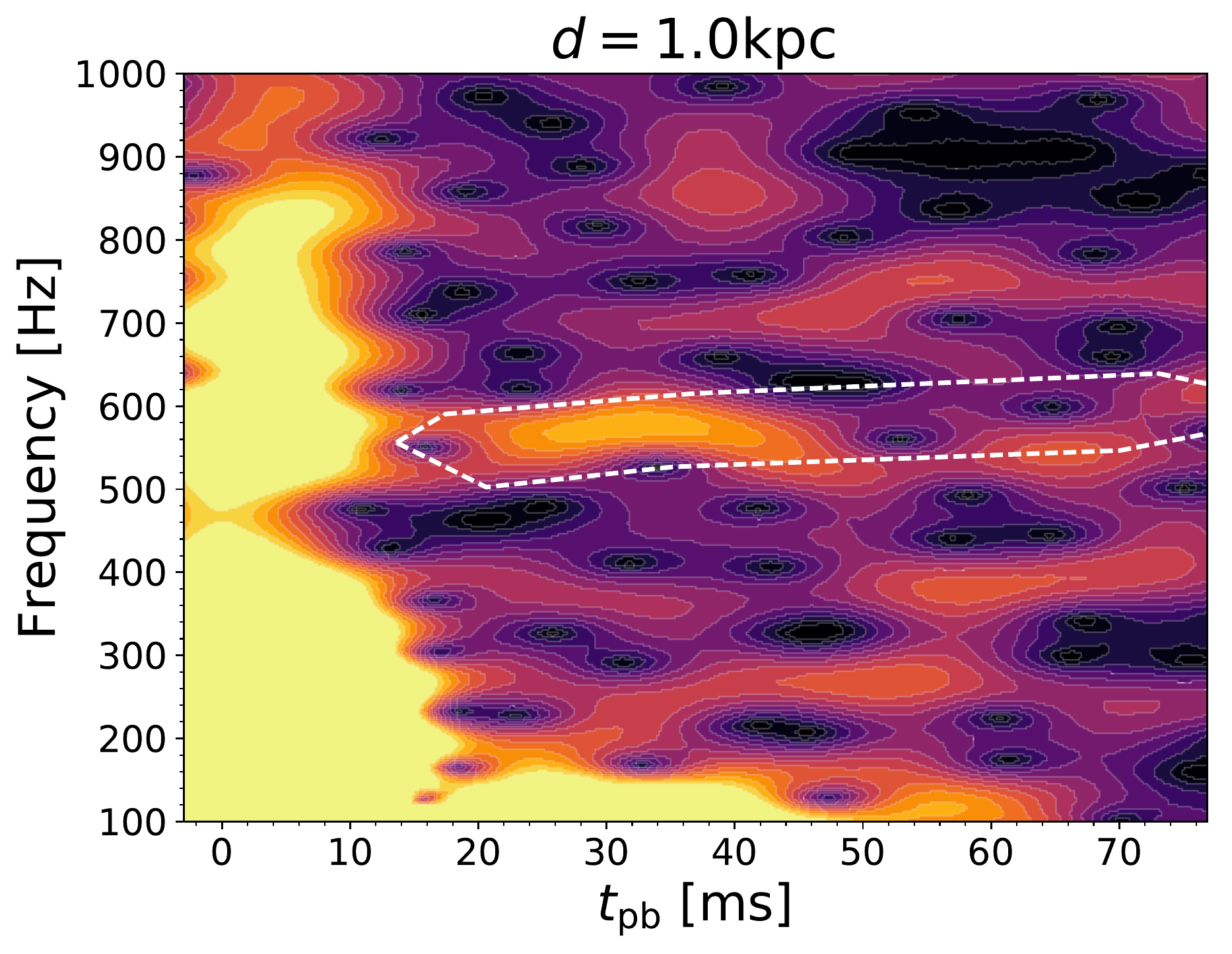}
\includegraphics[width=0.36\textwidth]{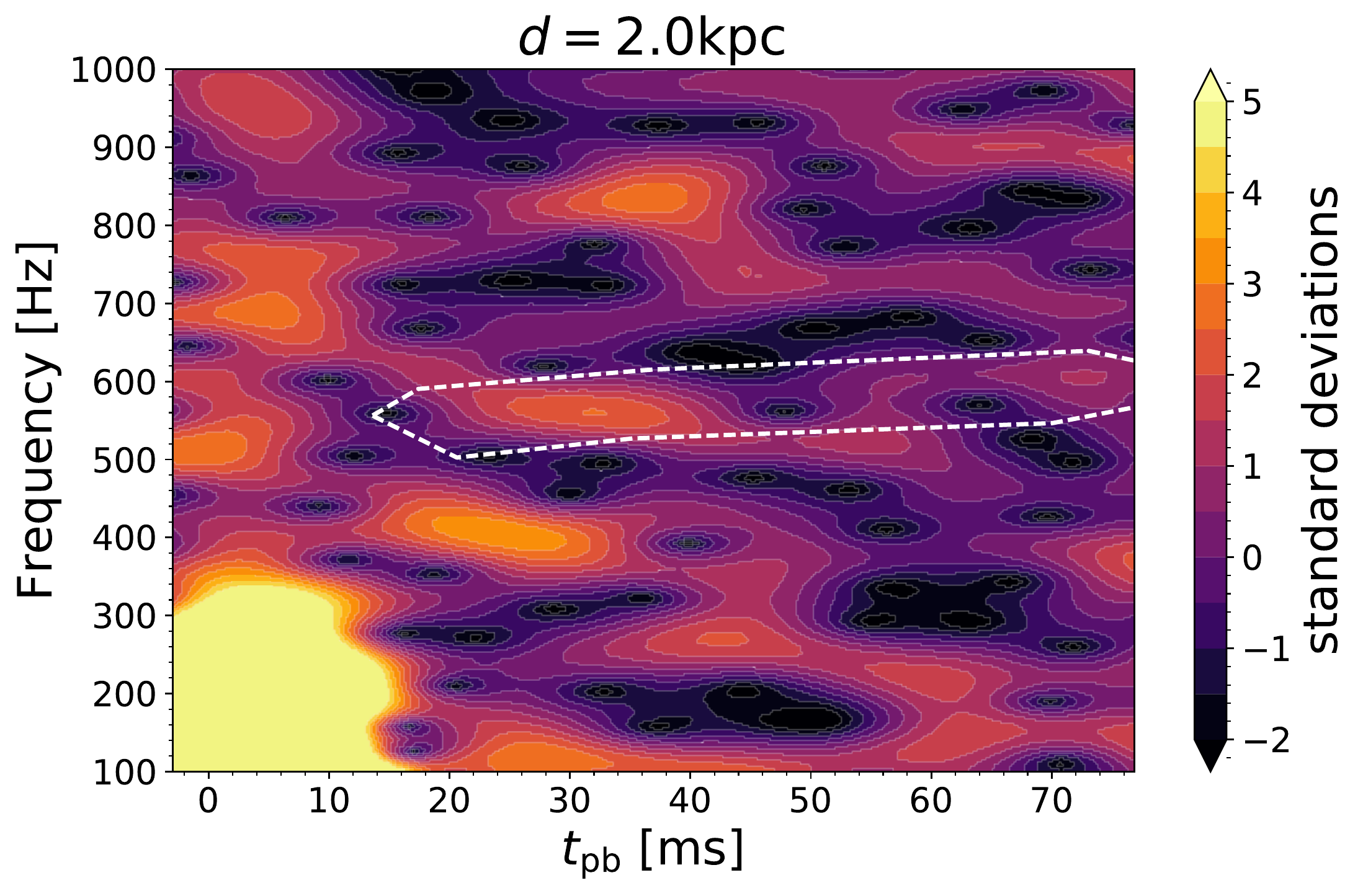}
\caption{Power spectral density maps of the expected IceCube signal for the $\Omega_c\!=\! 1.0$\,rad\,s$^{-1}$ model located at a distance of 0.5\,kpc (left), 1\,kpc (middle), and 2\,kpc (right). The data shown here correspond to the no oscillation scenario.  The colors denote standard deviations of pure Gaussian noise. See Supplemental Material at [see the published version] for animations of 128 realizations of the expected signal at each distance.} \label{fig:spectos}
\end{figure*}

In this section, we determine the distances at which the correlations we have found between the GW and neutrino signals are detectable with current detector technology.  For Advanced LIGO design sensitivity noise levels, we find the signal near 40\,ms after bounce in the $\Omega_c\! =\! 1.0$\,rad\,s$^{-1}$ simulation under optimal orientations should be observable out to a distance of $\sim$5\,kpc, while the imprint in the neutrino signal is observable in IceCube within a distance of $\sim$1\,kpc assuming the frequency is known from the gravitational wave signal.

\subsection{Gravitational Waves}

In \fref{fig:gwwave_train}, we showed a realization of the GW signal that would be seen by an equatorial observer detecting GWs from our $\Omega_c\! =\! 1.0$\,rad\,s$^{-1}$ simulation at a distance of 1\,kpc including a realization of Advanced LIGO design sensitivity noise. At 1\,kpc and comparable distances, the detection of the GW signal at the dominant frequencies we observe in our simulations should be possible. We now quantify this assertion, following \cite{Srivastava:2019}.  We use their Eq.~1 to estimate the distance at which a portion of our signal is observable (for an optimal orientation) at a signal-to-noise ratio of 8,
\begin{equation}
  d_{\mathrm{opt}} = \frac{\sigma}{\rho^*} = \frac{1}{\rho^*} \left[ 2 \int_{f_{\mathrm{low}}}^{f_{\mathrm{high}}} df \frac{\tilde{h}(f) \tilde{h}^*(f)}{S_h(f)}\right]^{1/2}\,,
\end{equation}
  where $\tilde{h}(f)$ is the Fourier transform of the $h_+(t)$ strain, $\tilde{h}^*(f)$ is the complex conjugate of this, and $S_h(f)$ is the design power spectral noise of Advanced LIGO. We take $f_{\mathrm{low}}=10$\,Hz and $f_{\mathrm{high}}=2048$\,Hz.  As in \cite{Srivastava:2019}, we take $\rho^*=8$, which is often taken to be the minimal signal-to-noise ratio for GW detections. Defining $h_+(t)$ at a distance of 1\,kpc gives $d_{\mathrm{opt}}$ in units of kpc.  We window our GW strain from this simulation using a Nuttall window function with a width of 40\,ms centered on 40\,ms after bounce.  With this narrowly defined time range, we find a value for $d_{\mathrm{opt}}$ of $\sim$5.5\,kpc, suggesting this signal is easily detectable at distances closer than this. As we shall see, this is much more promising than the neutrino prospects, and therefore we neglect a more detailed analysis and instead assume that we can obtain a clear indication of excited PNS mode frequencies with GWs (at distances closer than 5\,kpc) to aid our neutrino analysis.

\subsection{Neutrinos}
\label{sec:nudetect}
 In \sref{sec:sims} we also presented estimated IceCube rates for the $\Omega_c\!=\! 1.0$\,rad\,s$^{-1}$ simulation at 1\,kpc.  In this section, we generate realizations of IceCube event rates by adding detector noise (via the dark rate of the photomultiplier tubes (PMT), taken to be 550\,Hz per PMT \cite{IceCube2011AA}) and statistical noise from the finite neutrino arrival times (taken to be $\sqrt{N}$, where $N$ is the number of neutrinos expected within each 0.1\,ms time bin).  From these realizations we bin the mock data, window it using a Nuttall window with a 40\,ms width (although in practice the type of window does not impact the results), and Fourier transform the results to search for excess (and significant) power in time-frequency regions suggested by the GW signal as being potentially interesting.

In \fref{fig:spectos}, we show power spectral densities of the estimated IceCube neutrino rate as a function of time and frequency for several observer distances. These are similar to \fref{fig:rot10_l2mode_Lnusky}, but now for the expected detection rate rather than the luminosity of a specific neutrino species. We generate essentially random detector data by placing the source at a large distance. Fourier transforming $\sim 4\times10^7$ realizations of this data gives a flat (but noisy for any given realization) power spectrum for all frequencies greater than 100\,Hz with a characteristic median value set by the total number of events entering the windowed region and the number of bins, i.e. $P_{k} \sim N_{\mathrm{events}}/N_{\mathrm{bins}}^2$ \cite{lund2010}. We generate cumulative distributions of this noise power at 575\,Hz and at $\sim 40$\,ms after bounce (although this is arbitrary since the transform is dominated by noise) and determine the standard deviation levels which we use to normalize the data displayed in \fref{fig:spectos}. If the power is significantly above the median (defined as the $\sigma=0$ level), this is evidence for structure in the signal at that frequency and time. The broadband power at early times corresponds to the rapidly rising neutrino signal (see \fref{fig:rates}) at bounce. For close distances, $\sim$0.5\,kpc, the clear presence of the oscillations in the neutrino signal near $\sim$20-40\,ms between 500\,Hz and 600\,Hz is apparent in the Fourier transform. At $\sim$1\,kpc, the power spectral density still shows excess power at these times and frequencies, but its significance becomes weaker. It is not visible at 2\,kpc in this realization.

\begin{figure*}
\centering
\includegraphics[width=0.48\textwidth]{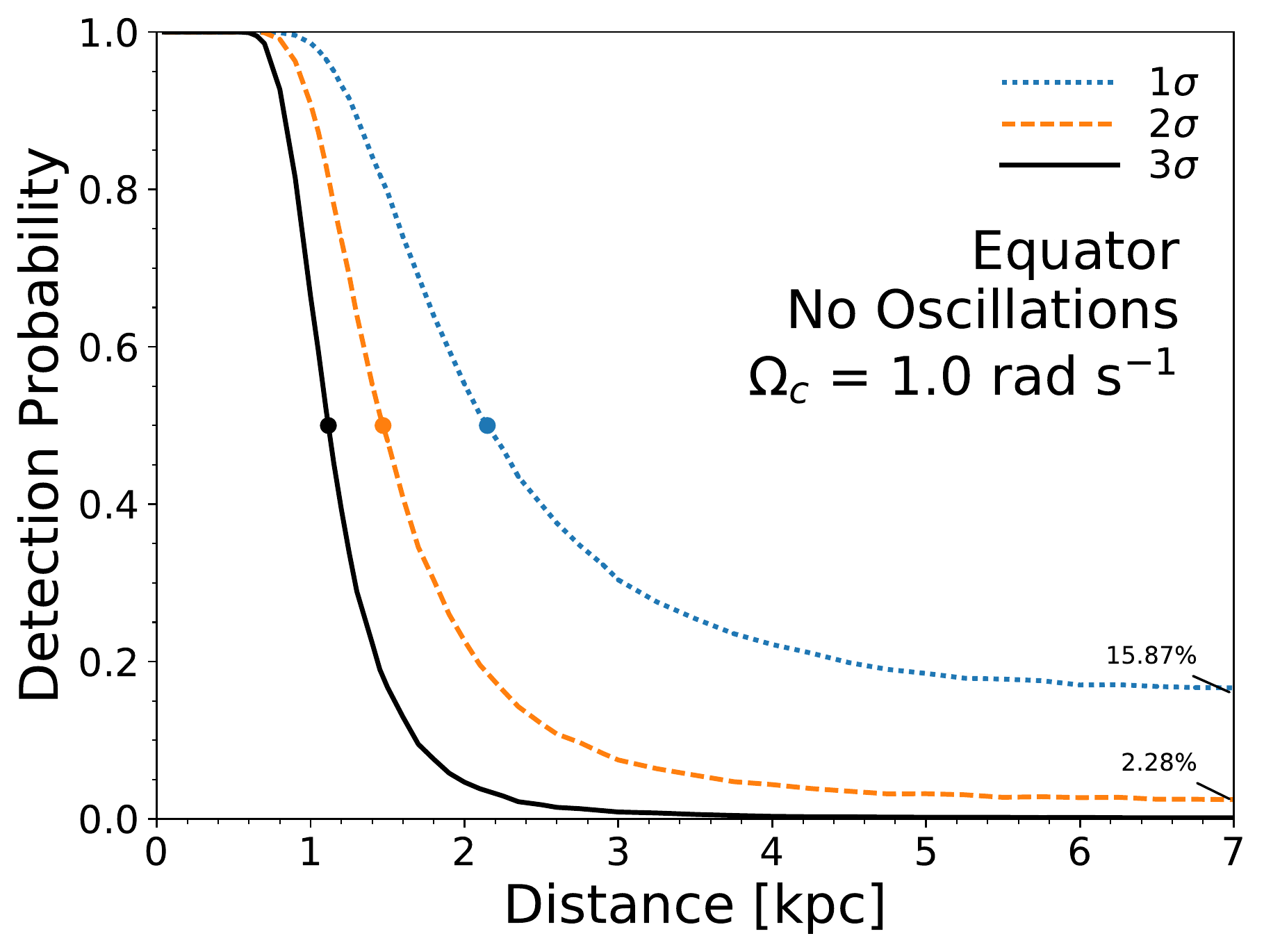}
\includegraphics[width=0.48\textwidth]{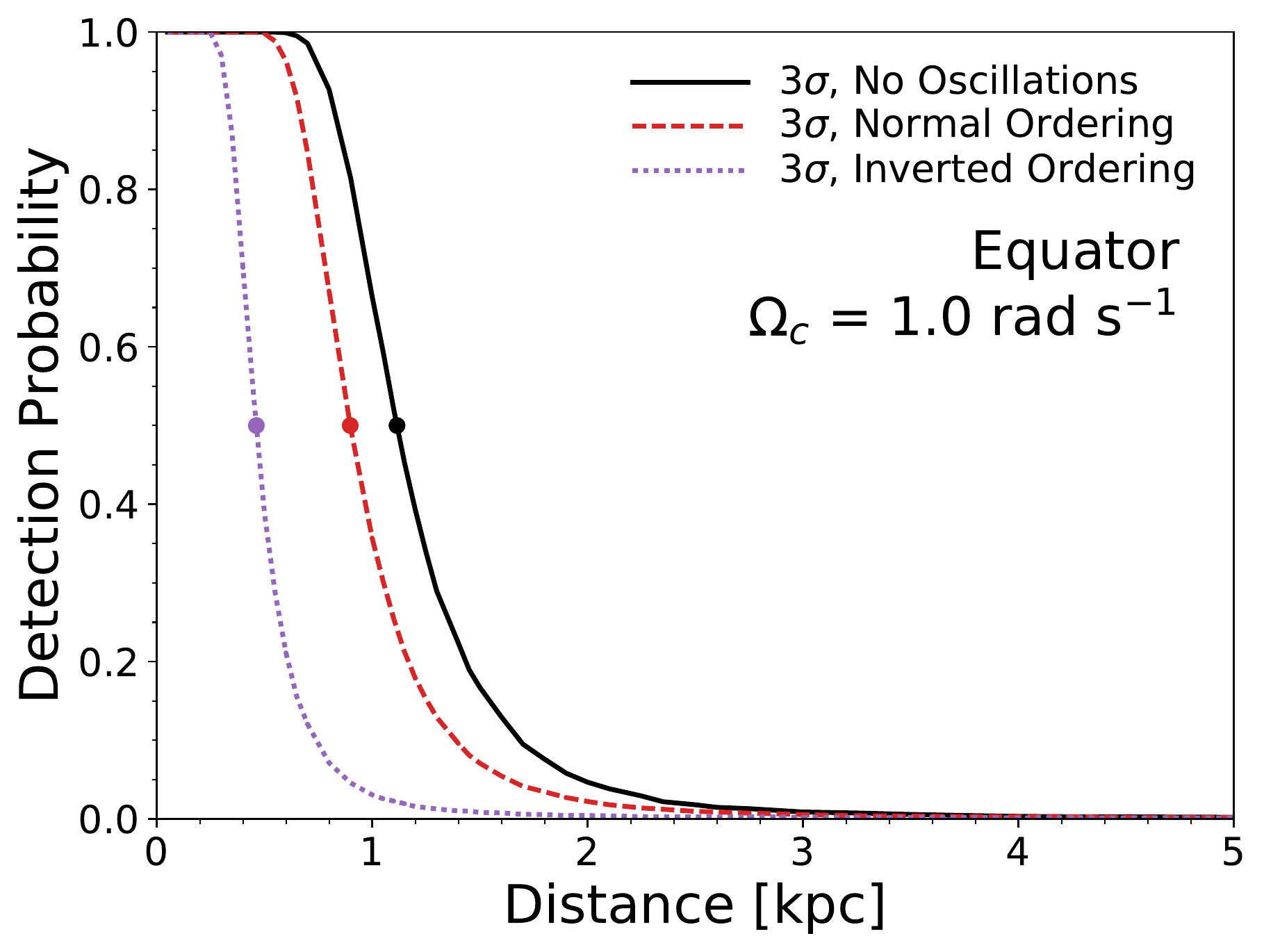}
\caption{Detection probability, defined as the fraction of realizations where the power spectral density at $t\!=\! 37$\,ms and $f\!=\! 575$\,Hz exceeds the 1$\sigma$, 2$\sigma$, and/or 3$\sigma$ levels defined by pure noise, vs distance for the $\Omega_c\!=\! 1.0$\,rad\,s$^{-1}$ simulation.  At close distances the oscillatory part is easily found, but quickly gets buried in the noise as the distance is increased.  In the left panel we show the 1$\sigma$ (blue dotted), 2$\sigma$ (orange dashed), and/or 3$\sigma$ (solid black) detection probabilities for an observer located on the equator and where the neutrinos underwent no oscillations after being emitted.  In the right panel we show the 3$\sigma$ detection probabilities for an observer located on the equator, but for three oscillation scenarios: no oscillations (solid black), normal ordering (dashed red), and inverted ordering (dotted purple).} \label{fig:detectability} 
\end{figure*}

To quantify the detectability, we determine the percentage chance of making at least a 1$\sigma$, 2$\sigma$, and 3$\sigma$ detection of excess power at $t\!=\! 37$\,ms and $f\!=\! 575$\,Hz for the $\Omega_c\!=\! 1.0$\,rad\,s$^{-1}$ simulation at varying distances.  We choose the specific point $t=37$\,ms and $f=575$\,Hz based on the expected GW detection at this time and frequency.  We construct these percentages by making 50000 realizations of the detected IceCube signal at each distance and determining the ratio of the realizations with a power of at least the 1$\sigma$, 2$\sigma$, and 3$\sigma$ level to the total number of realizations. We show these percentages in \fref{fig:detectability}.  In the left panel we show the 1$\sigma$, 2$\sigma$, and 3$\sigma$ detection percentage for the no oscillation scenario (and an equatorial observer). We note the chance of making at least a 1$\sigma$ (2$\sigma$, 3$\sigma$) detection of excess power from a pure noise signal is 15.87\% (2.28\%, 0.135\%), hence the asymptotic values at large distances.  For this panel, the distances the discovery potential for 1$\sigma$, 2$\sigma$, and 3$\sigma$ (defined as the probability of seeing a signal of this significance 50\% of the time) are $\sim$2.15\,kpc $\sim$1.47\,kpc, and $\sim$1.12\,kpc, respectively. In the right panel, we show the 3$\sigma$ detection percentage (also for an equatorial observer) for the three oscillation scenarios: no oscillations, normal ordering, and inverted ordering. For these scenarios, we predict the distances for a 3$\sigma$ discovery potential for the no oscillation, normal ordering, inverted ordering oscillation scenarios are $\sim$1.12\,kpc, $\sim$0.90\,kpc, and  $\sim$0.46\,kpc, respectively. The varying distances for the different oscillation scenarios reflect the different amplitudes of the oscillation signal in the $\bar{\nu}_e$ and $\nu_x$ signals.  As discussed for \fref{fig:rates}, the no oscillation signal is dominated by $\bar{\nu}_e$ while in the inverted ordering the signal is dominated by the $\nu_x$ signal.  The normal ordering is a mixture between $\bar{\nu}_e$ and $\nu_x$, but dominated by $\bar{\nu}_e$.

In Appendix~\ref{sec:modelneutrino}, we generalize the detectability of a small-amplitude, periodic signal on top of a constant background in the IceCube detector. Based on this toy model we derive a theoretical maximum distance for a detection (i.e. a 3$\sigma$ detection 50\% of the time) of,  

\begin{equation} 
  d^{3\sigma}_{\mathrm{th}} =1.57\,\mathrm{kpc} \left[\frac{\epsilon}{1}\right] \left[\frac{a}{0.01}\right] \left[\frac{A^{\mathrm{1\,kpc}}}{30000\,\mathrm{ms}^{-1}}\right]^{1/2} \left[\frac{\Delta \tau}{40\,\mathrm{ms}}\right]^{1/2}\,,\label{eq:nudetection}
\end{equation}

where $\epsilon$ is the purity of the signal (1 in the case of our toy model; 0.6-0.7 for our simulated signals; see Appendix~\ref{sec:modelneutrino}), $a$ is the fractional amplitude of the periodic signal, $\Delta \tau$ is the time frame over which the signal is present, and $A^{\mathrm{1\,kpc}}$ is the 1\,kpc-equivalent mean steady-state neutrino rate. For the latter, to be clear, $A^{\mathrm{1\,kpc}}$ is intrinsic to the source and not dependent on distance.  It is a function of the neutrino spectral properties through SNOwGLoBES.  This formula, and the detectability itself, is not a function of the frequency of the variation, as long as several cycles fall within the observing window $\Delta \tau$. This formula is valid for regimes where the signal is not overwhelmed by the detector background noise, for the conditions seen here, a few kpc (see \fref{fig:analytic} and the discussion in Appendix~\ref{sec:modelneutrino}).  This also means that even next-generation neutrino detectors, such as Hyper-Kamiokande \cite{abe18} and DUNE \cite{Abi:2018dnh}, will not be able to better measure this effect even though they are essentially background free.

As an application of this formula, we return to the oscillations observed in the neutrino signal for the $\Omega_c\!=\! 2.5$\,rad\,s$^{-1}$ simulation within 10\,ms after bounce.  There, $A^{\mathrm{1\,kpc}} \sim 10000\,\mathrm{ms}^{-1}$, $a \sim 0.04$, and $\Delta \tau \sim 5$\,ms.  This gives maximum detectable distances of $\sim$1\,kpc as well. In practice, the shorter window for which the oscillations are present as well as lower overall rate (and therefore stronger impact of the detector noise) may reduce this distance. We also note that this formula is not in disagreement with the estimate in \cite{lund2010}, where it is stated that a 1\% amplitude variation should be detectable at 10\,kpc.  This is because for that estimate $A^{\mathrm{1\,kpc}} \sim 135000\,\mathrm{ms}^{-1}$ and $\Delta \tau \sim 400$\,ms, giving a maximum detectable distance based on \eref{eq:nudetection} of $\sim$10.5\,kpc, as suggested in \cite{lund2010}.
 
\section{Outlook and conclusions} \label{ch:CCSNconc}

In this work, we simulated the core collapse and early post-bounce evolution of a $20\, M_\odot$ progenitor star with pre-collapse core rotations ranging from $0.0$-$2.5$~rad s$^{-1}$. We use axisymmetry for these simulations, which is a simplifying assumption, but justified for the collapse and early post-bounce phase for rotating stars given axisymmetric initial conditions, before turbulence in the gain region starts gaining dominance. For each simulation we extracted the GW and neutrino signals and showed these messengers can offer detailed asteroseismological information on the newly born PNS. These two messengers are complementary in that they carry information about certain linear modes of the core from different radii, with neutrinos probing the outer $60\!-\!80$\,km and GWs probing deeper in.

To characterize the modes, we followed a strategy of mode function matching, rather than mode frequency matching as in~\cite{torres2017towards,morozova2018gravitational, torres2018towards}. We believe this is a more robust approach that is less susceptible to mode misidentification, especially given the approximations employed in both simulations and perturbative schemes. By mode function matching we discovered that mode frequencies are overestimated by $\sim 15\%$ in our simulations. Our findings motivate further investigation to fully understand this mismatch in mode frequencies.

Many other spectral peaks exist in both GW and neutrino luminosity spectrograms along our entire sequence of rotating models, and we focused on a dominant peak. In~\cite{westernacher2018turbulence}, numerous additional modes were identified via mode function matching between perturbation theory and our non-rotating model, many of which are not quadrupolar. The many modes that are active offer to explain the additional spectral peaks in the multimessenger signals we observe along our rotating sequence.

The mechanism by which the linear modes of the core imprint themselves on the neutrino light curves appears to be that the neutrino-emitting volume (and possibly the local neutrino production rate) undergoes coherent deformations in time according to the frequency and angular harmonics of the active PNS modes in the vicinity of the neutrinospheres. The dominant angular harmonics are then reflected in the emission pattern of neutrinos on the sky at those frequencies. The comparison of the angular structure was made possible through the use of a grid-based two-moment transport scheme for the neutrinos because it retains and transports the directional emission information from the neutrinosphere region.

For the detection prospects, we focused on the $\Omega_c\!=\! 1.0$\,rad\,s$^{-1}$ simulation.  Using approximate assessment techniques, we determined that the imprint of the dominant mode of the GW signal is detectable within a distance of $\sim\! 5$\,kpc assuming the design sensitivity of Advanced LIGO.  Since the mode has a dual imprint, we have used the GW signal to inform a search for the same frequency in the neutrino signal, which we expect to be much more difficult to detect. This constitutes a bonafide multimessenger detection strategy, and allows us to assign a much higher significance than if no GW information was available, i.e. if we needed to search over many frequencies.

We then performed a detailed assessment of the detection prospects for the mode's neutrino imprint by looking at the expected signal in the IceCube Neutrino Observatory.  Given the amplitude of the mode's imprint is $\sim$1\% of the main neutrino signal, a detection requires very large events rates, and therefore a very close supernova, $\sim\! 1$\,kpc.  In the future, the
proposed IceCube-Gen2 will include twice the number of strings
compared to the current IceCube detector, which would increase the
number of detected neutrino events by a factor of 2 and increase the
range to detect this signal by a factor of $\sqrt{2}$. Further planned improvements in the
photosensors, which should allow for further
discrimination from the inherent background rate, is actively being studied by the IceCube
collaboration and could give rise to further improvements in the
detection distances mentioned here.

Lastly, the mechanism of the multimessenger imprint should generalize to other systems, eg. accretion-induced collapse of white dwarfs or binary neutron star post-merger remnants, although their distance makes detection in neutrinos unlikely. In these systems, the mode function matching procedure should also be useful for identifying the active modes in simulations, although with rapid rotation the perturbative schemes would have to be generalized beyond spherical symmetry.

\acknowledgements
We thank the anonymous referee for useful comments.
JRWS thanks Nick Stergioulas and John Friedman for sharing their insight on why the Cowling approximation systematically overestimates mode frequencies. JRWS also thanks Eric Poisson for first pointing out that the partially-relaxed Cowling approximation is not a controlled perturbative procedure. EOC thanks Peter Shawhan and Erik Katsavounidis
 for a discussion on LIGO detection prospects.  JRWS, EOC, EOS, IT, and MRW thank the Kavli Foundation, the Danish National Research Foundation (DNRF132), and the Niels Bohr institute in Copenhagen for their hospitality and for organizing the Kavli summer school in gravitational-wave astrophysics in 2017 where part of this work has been performed. The simulations were performed on resources provided by the Swedish National Infrastructure for Computing (SNIC) at PDC, HPC2N, and NSC.  JRWS acknowledges support from OGS. EOC acknowledges support from the Swedish Research Council (Project No.~2018-04575). IT thanks the Villum Foundation (Project No.~13164), the Knud H\o jgaard Foundation, and the  Deutsche Forschungsgemeinschaft through 
Sonderforschungsbereich SFB~1258 ``Neutrinos and Dark Matter
in Astro- and Particle Physics (NDM).
MRW acknowledges support from the Ministry of Science and Technology, Taiwan under Grant No. 107-2119-M-001-038, and
the Physics Division, National Center of Theoretical Science of Taiwan. SMC is supported by the U.S. Department of Energy, Office of Science, Office of Nuclear Physics, under Award numbers DE-SC0015904 and DE-SC0017955.

Software: PyCBC \cite{pycbc}, Matplotlib \cite{Hunter:2007}, FLASH \cite{fryxell2000flash,dubey2009extensible,couch13, o2018two}, NuLib \cite{o2015open}, SciPy \cite{scipy}, SNOwGLoBES
\cite{scholberg:12}.

%
%
\appendix
\section{Tests of Perturbative Schemes} \label{sec:CCSN_test_pert_schemes}

In~\cite{westernacher2018turbulence} the perturbative schemes of~\cite{torres2017towards} and~\cite{morozova2018gravitational} were tested on a stable TOV star with polytropic equation of state $P = K \rho^\Gamma$ with $\Gamma=2$, $K=100$, and central density $\rho_c = 1.28 \times 10^{-3}$ in geometrized units. The purpose of testing on this compact star is to show that the regime of validity of partially-relaxed Cowling approximations deserves independent investigation, and that the \texttt{FLASH} implementation tends to overestimate mode frequencies with respect to the full Cowling approximation. The scheme of~\cite{morozova2018gravitational} is to allow the lapse function to vary, but all other metric functions are fixed. This scheme is not \emph{a priori} under control, since not all terms are accounted for at a given order. Indeed, a further relaxation of the Cowling approximation in~\cite{torres2018towards} resulted in corrections of a similar size as those obtained when going from a fixed spacetime to a varying lapse function. It was shown in~\cite{westernacher2018turbulence} for a TOV star that the partially-relaxed Cowling approximation results in worse determinations of fundamental mode frequencies than in the full Cowling approximation, and the radial order $n$ of mode functions is captured increasingly inaccurately for increasing $n$. The partially-relaxed Cowling approximation was not tested in~\cite{morozova2018gravitational}, nor in a subsequent study~\cite{radice2019characterizing}.

We reproduce the main results of these tests in Tables~\eqref{tab:TOVCowling} and ~\eqref{tab:TOVconformalflat}. More details are provided in~\cite{westernacher2018turbulence}.

\subsection{TOV Mode Test in the \texttt{FLASH} Implementation} \label{sec:FLASHTOV_test}

In this study we use the \texttt{FLASH}~\cite{fryxell2000flash,dubey2009extensible} implementation of~\cite{o2018two}, which uses Newtonian hydrodynamics and a phenomenological effective gravitational potential developed in~\cite{Keil1997,rampp2002radiation,marek2006exploring}, designed to mimic general relativity in spherical symmetry. The Newtonian hydrodynamics and effective gravitational treatment affect the mode frequencies obtained in simulations. In~\cite{westernacher2018turbulence} the modes of a stable TOV star  were extracted in our \texttt{FLASH} implementation. This test is relevant to our study since the dominant modes of oscillation are extracted from CCSN simulations within the \texttt{FLASH} implementation.

TOV \emph{migration tests} was carried out in~\cite{marek2006exploring,o2018two}, where a TOV star on the unstable branch is observed to migrate to the stable branch. Note these TOV solutions are computed using the equations that correspond to the pseudo-Newtonian system (i.e. case A of \cite{marek2006exploring}), and therefore are equilibrium configurations in FLASH. The ensuing oscillations were observed to have a frequency $\sim 2$x higher than in the general relativistic case. In~\cite{westernacher2018turbulence} the fundamental radial ($l\!=\! 0$) and axisymmetric quadrupolar ($l\!=\! 2$, $m\!=\! 0$) modes $\lbrace F$, ${}^2 f \rbrace$ and their overtones $\lbrace H_1$, $H_2$, ${}^2 p_1$, ${}^2 p_2$, ${}^2 p_3 \rbrace$ were extracted from the same stable TOV star studied in \sref{sec:CCSN_test_pert_schemes}. The main results are reproduced from~\cite{westernacher2018turbulence} in Table~\eqref{tab:FLASHTOV_mode_test}.

The main conclusion of this test is that, except for the fundamental radial mode, our \texttt{FLASH} implementation is overestimating mode frequencies even with respect to the Cowling approximation. Since the Cowling approximation itself overestimates frequencies, we can conclude that \emph{the true frequencies of modes we identify in our CCSN simulations are bounded above by the frequency corresponding to the best-fit mode functions}. From the perspective of the Nyquist frequency, this is a favorable conclusion for detection prospects, since lower frequencies can be resolved with a lower event rate. However, in practice the detection threshold is far from the Nyquist limit, becoming independent of frequency for fixed signal duration (changes in frequency result in compensating changes in the total number of periods present over the signal duration), see \sref{sec:detect}. Thus the dominant variables for detection prospects are instead the signal amplitude and duration.

\section{Model Neutrino Detection Distances}
\label{sec:modelneutrino}

\begin{table*}
\centering
\scalebox{1.0}{%
\begin{tabular}{l|l|l|l|l|l|l|l|l|l|l|l|l|}

 & ${}^1 f$ & ${}^1 p_1$ & ${}^1 p_2$ & ${}^1 p_3$ & ${}^2 f$ & ${}^2 p_1$ & ${}^2 p_2$ & ${}^2 p_3$ & ${}^3 f$ & ${}^3 p_1$ & ${}^3 p_2$ & ${}^3 p_3$ \\ \hline
From~\cite{font2001axisymmetric} (kHz)                                                                                   & 1.335    & 3.473      & 5.335      & 7.136      & 1.846    & 4.100      & 6.019      & 7.867      & 2.228    & 4.622      & 6.635      & 8.600      \\ \hline
From~\cite{gaertig2008oscillations} (kHz)                                                                                & -        & -          & -          & -          & 1.890    & 4.130      & -          & -          & -        & -          & -          & -          \\ \hline
Current work (kHz)                                                                                                                    & 1.376 & 3.469 & 5.336 & 7.141 & 1.881 & 4.104 & 6.028 & 7.866           & 2.255 & 4.640 & 6.647 & 8.535           \\ \hline \hline
\% diff. with~\cite{font2001axisymmetric}    & 3.7         & 0.12           & 0.019 & 0.070 & 2.4 & 0.096 & 0.15 & 0.013 & 1.2 & 0.39 & 0.18 & 0.76           \\ \hline
\% diff. with~\cite{gaertig2008oscillations} & -        & -          & -          & -          & 0.48 & 0.63 & -          & -          & -        & -          & -          & -         
\end{tabular}}
\caption{A comparison between the mode frequencies we obtain in the Cowling approximation with a boundary condition of vanishing radial displacement at the stellar surface, and those obtained in~\cite{font2001axisymmetric,gaertig2008oscillations} using different methods, for a $\Gamma =2$, $K=100$, $\rho_{0,c}=1.28\times 10^{-3}$ TOV star (in geometrized units). The dominant error in the frequency is in the specification of location of the stellar surface, at which the boundary condition is imposed; changing it by one grid point yields a possible modification of the frequencies by $\sim 1$ Hz.} \label{tab:TOVCowling}

\centering
\begin{tabular}{l|l|l|l|l|}

                         & ${}^2 f$ & ${}^2 p_1$ & ${}^4 f$ & ${}^4 p_1$ \\ \hline
From~\cite{dimmelmeier2006non}, GR CFC (kHz) & 1.586    & 3.726      & 2.440    & 4.896      \\ \hline
Current work, partial Cowling (kHz)       & 2.496    & 3.777      & 3.047    & 4.999      \\ \hline
Current work, Cowling (kHz) & 1.881 & 4.104 & 2.565 & 5.112      \\ \hline \hline
\% diff.~\cite{dimmelmeier2006non} vs partial Cowling                 & 57      & 1.4        & 25     & 2.1 \\ \hline
\% diff.~\cite{dimmelmeier2006non} vs Cowling & 19 & 10 & 5.1 & 4.4 \\ \hline
\end{tabular}
\caption{A comparison between the mode frequencies we obtain perturbatively using the partially-relaxed Cowling approximation of~\cite{morozova2018gravitational} and those obtained in~\cite{dimmelmeier2006non} using full numerical simulations in the conformal flatness approximation, for the same $\Gamma \!=\!2$, $K\!=\!100$, $\rho_{0,c}\!=\!1.28\times\! 10^{-3}$ TOV star. The conformal flatness approximation is regarded as quite accurate for these modes~\cite{friedman2013rotating}. The agreement with~\cite{dimmelmeier2006non} is worsened considerably for the fundamental modes, but improved for the overtones shown, in comparison to the frequencies obtained in the full Cowling approximation. It has been observed almost always that the Cowling approximation tends to overestimate the true frequencies, see eg.~\cite{font2002three,cerda2008new,zink2010frequency, chirenti2015fundamental,mendes2018new}. However, see the fundamental radial mode appearing in Fig.~11 of~\cite{torres2018towards} for an apparently glaring exception.}
\label{tab:TOVconformalflat}
\centering
\begin{tabular}{l|l|l|l|l|l|l|l|}

                         & $F$ & $H_1$ & $H_2$ & ${}^2 f$ & ${}^2 p_1$ & ${}^2 p_2$ & ${}^2 p_3$ \\ \hline
From~\cite{dimmelmeier2006non} \&~\cite{font2002three}, GR CFC (kHz) & 1.442    & 3.955  & 5.916      & 1.586    & 3.726 & - & -     \\ \hline
Current work, Cowling (kHz)       & 2.696    & 4.534 & 6.346     & 1.881    & 4.104  &  6.028 &  7.866    \\ \hline
Current work, \texttt{FLASH} (kHz) & 2.174 & 5.522 & 8.295 & 2.024 & 5.122 & 7.920 & 10.593      \\ \hline \hline
\% diff. \texttt{FLASH} vs GR CFC                 & +51      & +40        & +40     & +28 & +37 & - & - \\ \hline
\% diff. \texttt{FLASH} vs Cowling & -19 & +22 & +31 & +8 & +25 & +31 & +35 \\ \hline
\end{tabular}
\caption{A comparison between the mode frequencies we obtain from \texttt{FLASH} simulations and those obtained in the Cowling approximation and in full GR in the conformal flatness approximation (GR CFC), for the same $\Gamma\!=\! 2$, $K\!=\! 100$, $\rho_{0,c}\!=\! 1.28\times 10^{-3}$ TOV star. The \texttt{FLASH} simulations yield frequencies overestimated with respect to full GR in all cases. We observe an improvement in the fundamental radial mode frequency with respect to the Cowling approximation (i.e.~a downward correction), whereas all other mode frequencies obtain an erroneous upward correction.}
\label{tab:FLASHTOV_mode_test}
\end{table*}

The varying detectability distances in the different ordering scenerios seen in \fref{fig:detectability} stem from the different amplitudes of the oscillations in the $\bar{\nu}_e$ ($\sim\pm 1$\%) and $\nu_x$ ($\sim \pm 0.5$\%).  In order to generalize the determination of the detectability of arbitrary amplitude signals embedded in IceCube CCSN data we adopt a simple model.  The model is a flat and steady-state detection rate with a magnitude of $A^{\mathrm{1\,kpc}}$ at 1\,kpc, an oscillatory component with relative amplitude $a$, and frequency $f$.  For the fiducial distance of 1\,kpc, we take both $A\!=\! 30000$\,ms$^{-1}$ and $A\!=\! 60000$\,ms$^{-1}$, these are similar to the rates seen in \fref{fig:rates}.   We take $f\!=\! 300$\,Hz and 600\,Hz and vary $a$ from $10^{-3}$ to $10^{-0.5}$.  We also explore varying the Fourier transform window function width, $\Delta \tau$.

\begin{figure}
\centering
\includegraphics[width=0.48\textwidth]{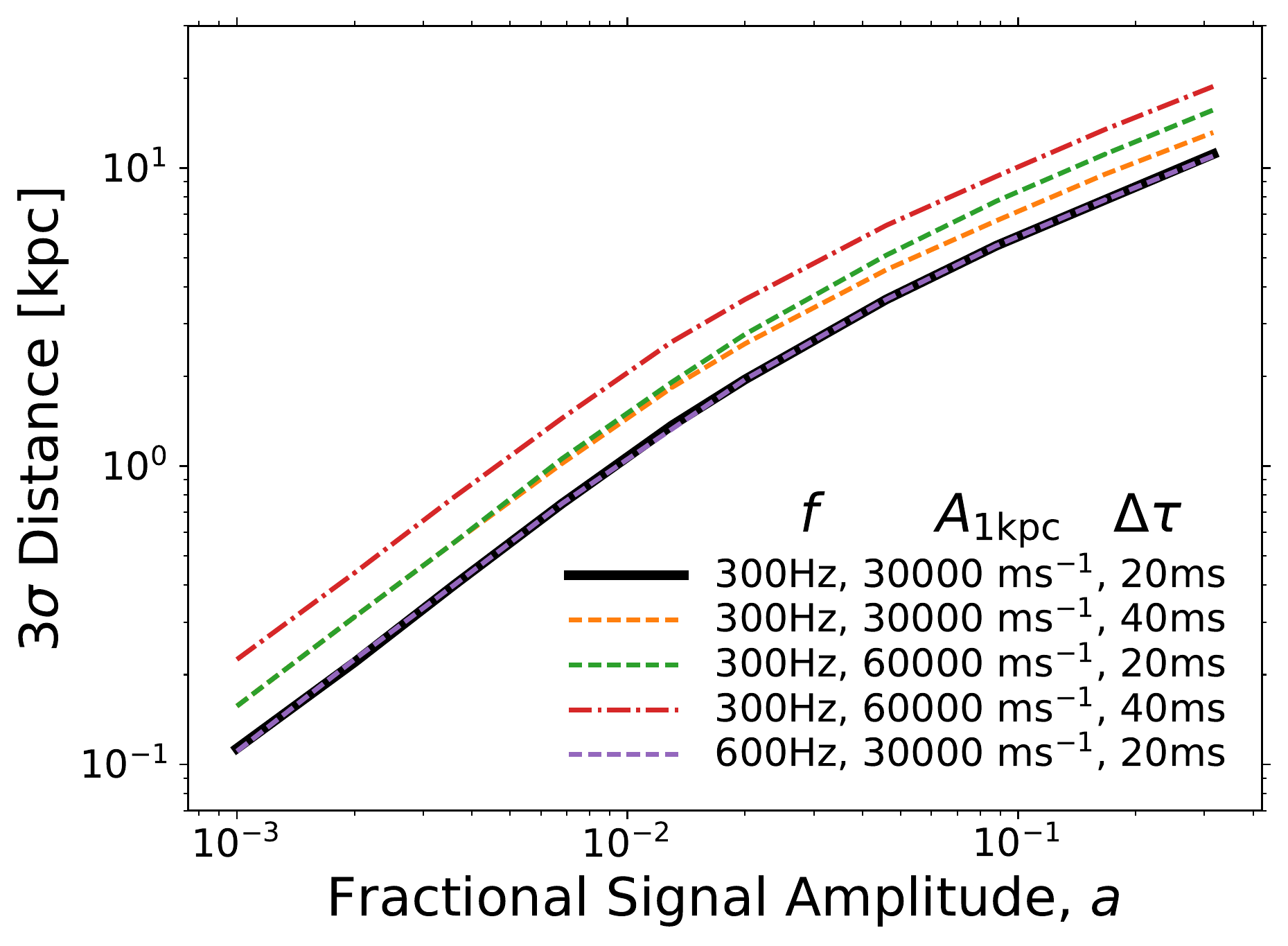}
\caption{Theoretical distances where a 3$\sigma$ detection of excess power at a particular frequency, $f$, would happen 50\% of the time.  The underlying signals that we analysed are from sample realizations of a flat background (with an amplitude of $A^{\mathrm{1\,kpc}}$ at 10\,kpc and scaled using the inverse square law for other distances) plus an oscillatory signal at frequency $f$ with a fractional amplitude of $a$. Detector noise from the PMTs and statistical counting noise is included in the realization as well.} \label{fig:analytic}
\end{figure}

Using the same technique as above, for each $A, a, f$, and $\Delta \tau$, we determine the maximum distance at which we would recover a 2$\sigma$ or 3$\sigma$ discovery potential. Note, we keep the intrinsic CCSN luminosity fixed, therefore we adjust $A$ with distance following the inverse square law, i.e. $A\!=\! 30000/\mathrm{ms} (1\,\mathrm{kpc}/r)^2$. We show the results in \fref{fig:analytic}, where we plot the distances at which we achieve a 3$\sigma$ discovery potential for a given fractional signal amplitude $a$. At closer distances the fraction of 3$\sigma$ detections quickly increases, as seen in \fref{fig:detectability}. Generally, we need close by sources (where the detector signal is high) in order to identify small amplitude signals, while large fractional amplitudes can be detected out to much larger distances. At nearby distances ($d<2$\,-\,$3$\,kpc) the dark noise rate in the detector does not impact the maximum detectable distance. For larger distances, where the rate in the detector is small, the detector photomultiplier noise inhibits this detection.

We make the following observations.  Doubling the window over which we search for a given frequency (or taken another way, the length over which an oscillatory signal is present in the data) increases the maximum distance by a factor $\sqrt{2}$ (solid black $\to$ dashed orange) as does doubling in the intrinsic rate (solid black $\to$ dashed green).  Of these two, doubling the rate gives larger maximum distances for the largest amplitudes because the smaller window size limits the impact of the detector photomultiplier noise.  Doubling both the intrinsic rate and the window leads to a factor of 2 increase in the maximum distance (solid black $\to$ red dashed). Lastly, doubling the frequency of the oscillatory mode, while keeping the fractional amplitude and time range over which it is present fixed, has no impact on the maximum distance (solid black $\to$ dashed purple). These curves are consistent with our results in \sref{sec:nudetect} and motivate \eref{eq:nudetection}, repeated here for completeness,

\begin{equation} 
  d^{3\sigma}_{\mathrm{th}} =1.57\,\mathrm{kpc} \left[\frac{\epsilon}{1}\right] \left[\frac{a}{0.01}\right] \left[\frac{A^{\mathrm{1\,kpc}}}{30000\,\mathrm{ms}^{-1}}\right]^{1/2} \left[\frac{\Delta \tau}{40\,\mathrm{ms}}\right]^{1/2}\,.
\end{equation}

We note that  for the $\Omega_c\!=\! 1.0$\,rad\,s$^{-1}$ case explored above, with an observer positioned on the equatorial plane, $A^{\mathrm{1\,kpc}}=30000$\,ms$^{-1}$, $\Delta \tau \sim 40$\,ms, and a fractional amplitude of $a=1\%$ (for no oscillations) and $a=0.5\%$ (for inverted ordering), these results suggest a maximum distance to which these oscillation are detectable of 1.57\,kpc and 0.79\,kpc, respectively.  These compare to our actual distances determined above of 1.12\,kpc and 0.46\,kpc for the no oscillation scenario and the inverted ordering, respectively.  The discrepancy that is present sets the purity (with values around 0.6-0.7) since the simulated signals are not pure oscillatory stationary signals on a flat background.

\section{Dependence of mode identification on boundary conditions}
\label{sec:modedurability}

In this section, we show that the mode identified via mode function matching in \sref{sec:modeextract} does not depend sensitively on different choices of boundary conditions, aside from the existence of an additional node at large radius ($r\sim 120\,$km) when placing the outer boundary condition at the shockwave location. The number of nodes more clearly within the PNS is 2 in all cases. The main reason for this independence is that our mode function matching is performed on a density-weighted basis. The different boundary conditions produce variations in the mode function morphology primarily at larger radii, which is suppressed by the $\sqrt{\rho}$-weighting we use for the matching procedure. Despite this suppression, the mode functions in each spectrum are distinct enough at smaller radii to allow for a convincing match with the simulation data.

We consider 4 boundary conditions, all using the full Cowling approximation. Firstly, the outer boundary condition can be imposed at different radii. In this work we imposed it at the shockwave location, which we take to be where the radial derivative of the spherically-averaged radial velocity is maximally negative, whereas in~\cite{morozova2018gravitational} it was imposed at the approximate location of the PNS surface (where $\rho = 10^{10}\,$g$\,$cm$^{-3}$). In our case, the PNS surface is not well-defined because we focus on a much earlier post-bounce phase than~\cite{morozova2018gravitational}. Another choice to make is whether to impose the vanishing of the radial displacement, $\eta_r |_{\mathrm{boundary}} =0$, as done in~\cite{torres2017towards} and in this work, or to impose the vanishing of the Lagrangian pressure perturbation, $\Delta P \vert_{\mathrm{boundary}} = 0$, as done in~\cite{morozova2018gravitational}. The latter corresponds to a free surface.

\begin{figure*}[htbp]
\centering
\includegraphics[width=\textwidth]{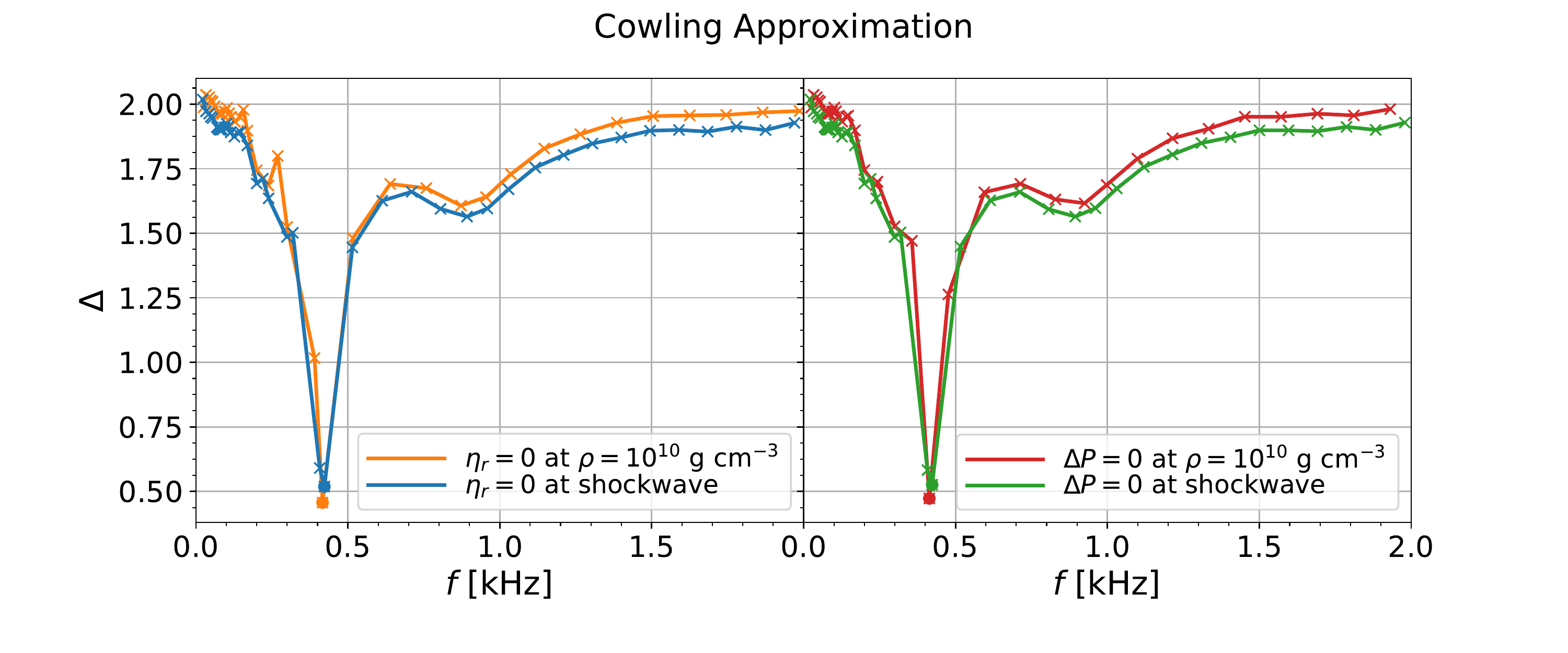}
\caption{Modefunction mismatch $\Delta$ (see \eref{eq:Delta}) between the perturbative mode spectrum and the simulation data using different boundary conditions. The simulation data being matched corresponds to the non-rotating model around 40 ms after bounce. Mode frequencies are indicated with crosses, and the best-fit mode frequency is indicated with a circle.} \label{fig:fitquality}
\end{figure*}

\begin{figure*}[htbp]
\centering
\includegraphics[width=\textwidth]{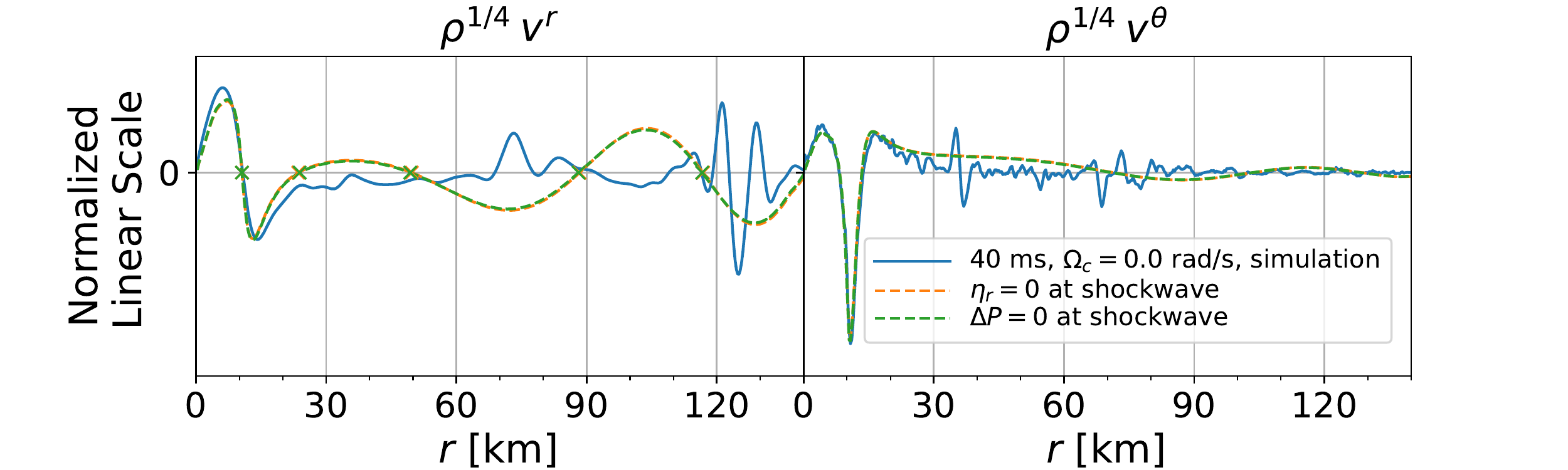}
\caption{Second best modefunction match in the Cowling approximation when the outer boundary condition is placed at the shockwave.} \label{fig:nextnearest}
\end{figure*}

In \fref{fig:fitquality} we plot the mismatch $\Delta$ (see \eref{eq:Delta}) between the perturbative mode spectrum and simulation data for the 4 boundary conditions just mentioned. The simulation data being matched is a snapshot from the non-rotating model around $40$ ms. Mode frequencies are indicated with crosses, and the best-fit mode frequency is indicated with a circle. The frequency of the best-fit mode functions all cluster around $420$ Hz. When imposing the boundary condition at $\rho=10^{10}\,$g$\,$cm$^{-3}$, there are no mode functions that compete with the quality of fit of the best-fitting one. When the boundary condition is at the shockwave, there appears to be one mode function around $405$ Hz with a similar quality fit. This competing mode function has an additional node at $r\sim 24$ km (see \fref{fig:nextnearest}), whereas the simulation data does not indicate zero-crossing behavior there. We therefore reject that mode by inspection.

\begin{figure*}[htbp]
\centering
\includegraphics[width=\textwidth]{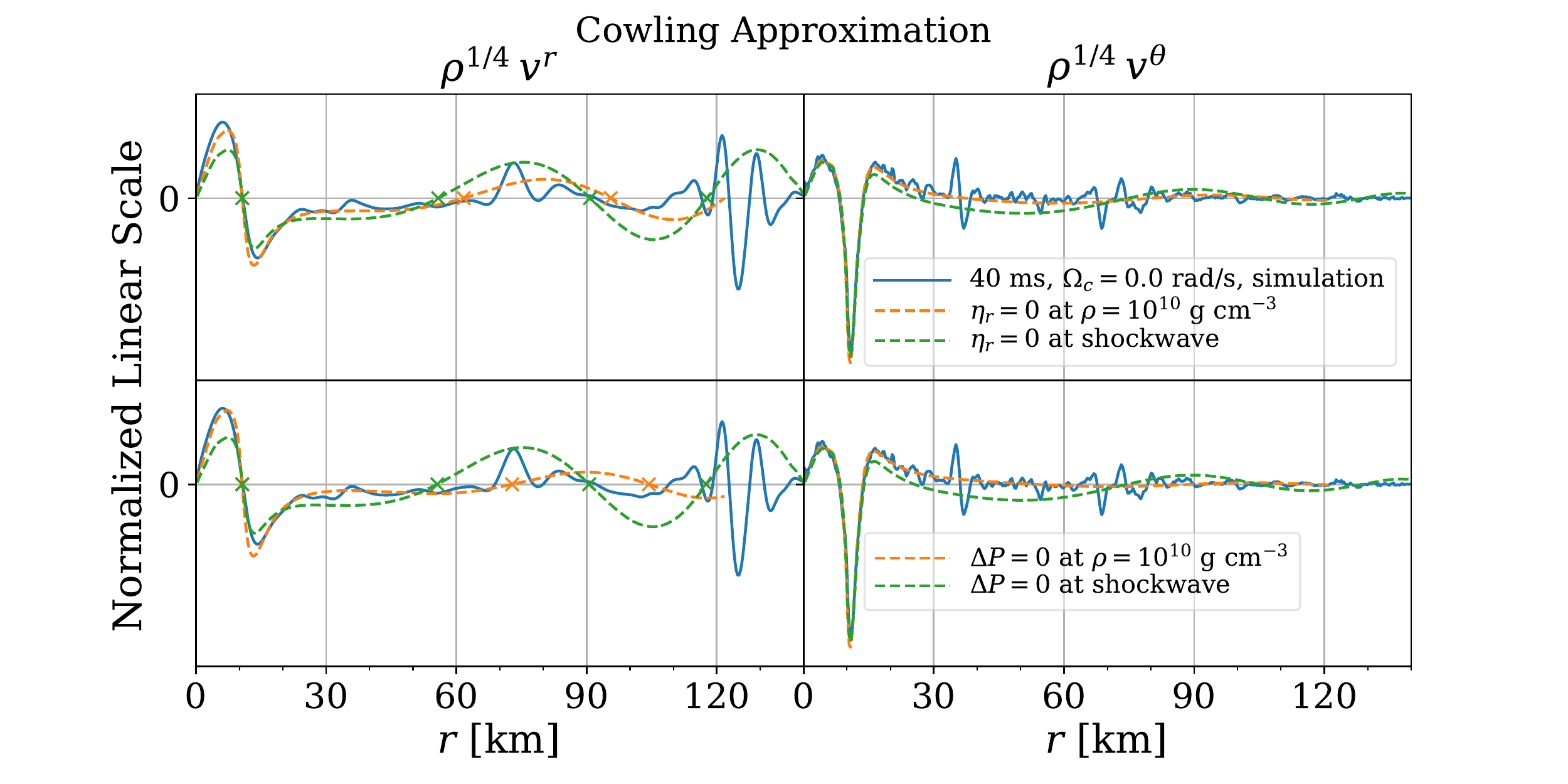}
\caption{Best-fit modefunctions from \fref{fig:fitquality}. Radial nodes are indicated with crosses. Compared with imposing the outer boundary condition at $\rho=10^{10}\,$g$\,$cm$^{-3}$, imposing it at the shockwave results in one additional node near $r=120\,$km. Otherwise, the modefunction morphologies are similar.} \label{fig:modefunc_diffbcs}
\end{figure*}

We also compare the best-fit mode functions for all 4 boundary conditions in \fref{fig:modefunc_diffbcs}. The density weighting used for these plots is $\rho^{1/4}$, which is weaker than the $\sqrt{\rho}$-weighting used for the modefunction matching and allows for easier visual inspection of the zero-crossing behavior at large radius. Compared to imposing the outer boundary condition at $\rho=10^{10}\,$g$\,$cm$^{-3}$, when imposing the outer boundary condition at the shockwave there is an additional node at large radius ($r\sim 120\,$km). Otherwise, the mode function morphologies are similar. 

Notice that the boundary condition yielding the best fit seems to be $\eta_r \vert_{\mathrm{boundary}}$ at $\rho=10^{10}\,$g$\,$cm$^{-3}$. However, one should not read too much into this, since the perturbative scheme being applied is not the consistent linearization of the simulated equations. The main observation we make here is the degree of independence of choice of boundary condition. We refrain from inferring which boundary condition is more correct from these comparisons. By contrast, in~\cite{sotani2019dependence} boundary conditions were assessed in this manner, even though the perturbative scheme is not consistent with their full-GR simulations (and one actually expects systematic over-estimation of mode frequencies by their Cowling perturbative spectra).

\subsection{Cowling versus partial Cowling approximations}

\begin{figure*}[htbp]
\centering
\includegraphics[width=\textwidth]{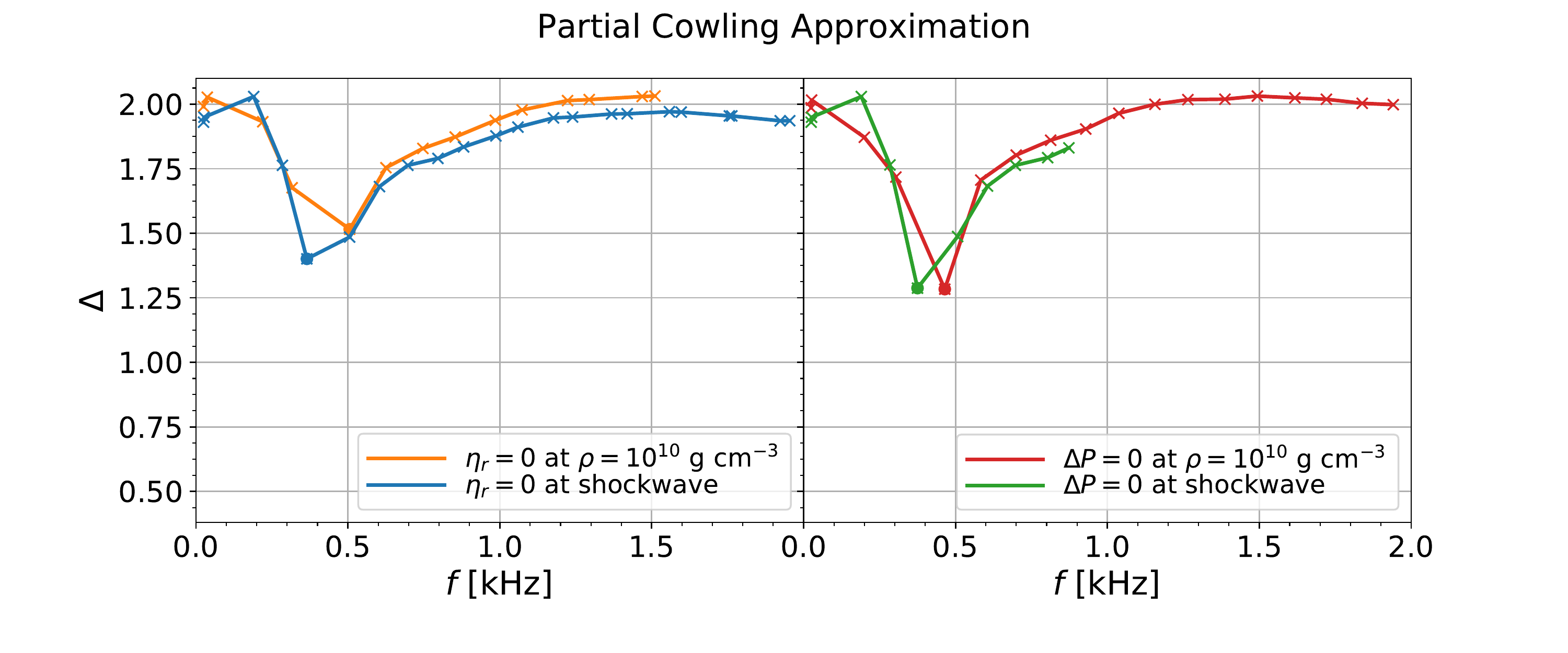}
\caption{Same as \fref{fig:fitquality} except using the partial Cowling approximation of~\cite{morozova2018gravitational}. For the boundary condition used in~\cite{morozova2018gravitational} ($\Delta P = 0$ where $\rho = 10^{10}\,$g$\,$cm$^{-3}$), the best-fit mode function has a frequency of $465$ Hz, quite close to our simulation value of $490$ Hz. However, the best-fit mode function in that case is a poor fit compared to the Cowling mode functions, and exhibits one additional node. } \label{fig:fitquality_partialCowling}
\end{figure*}

\begin{figure*}[htbp]
\centering
\includegraphics[width=\textwidth]{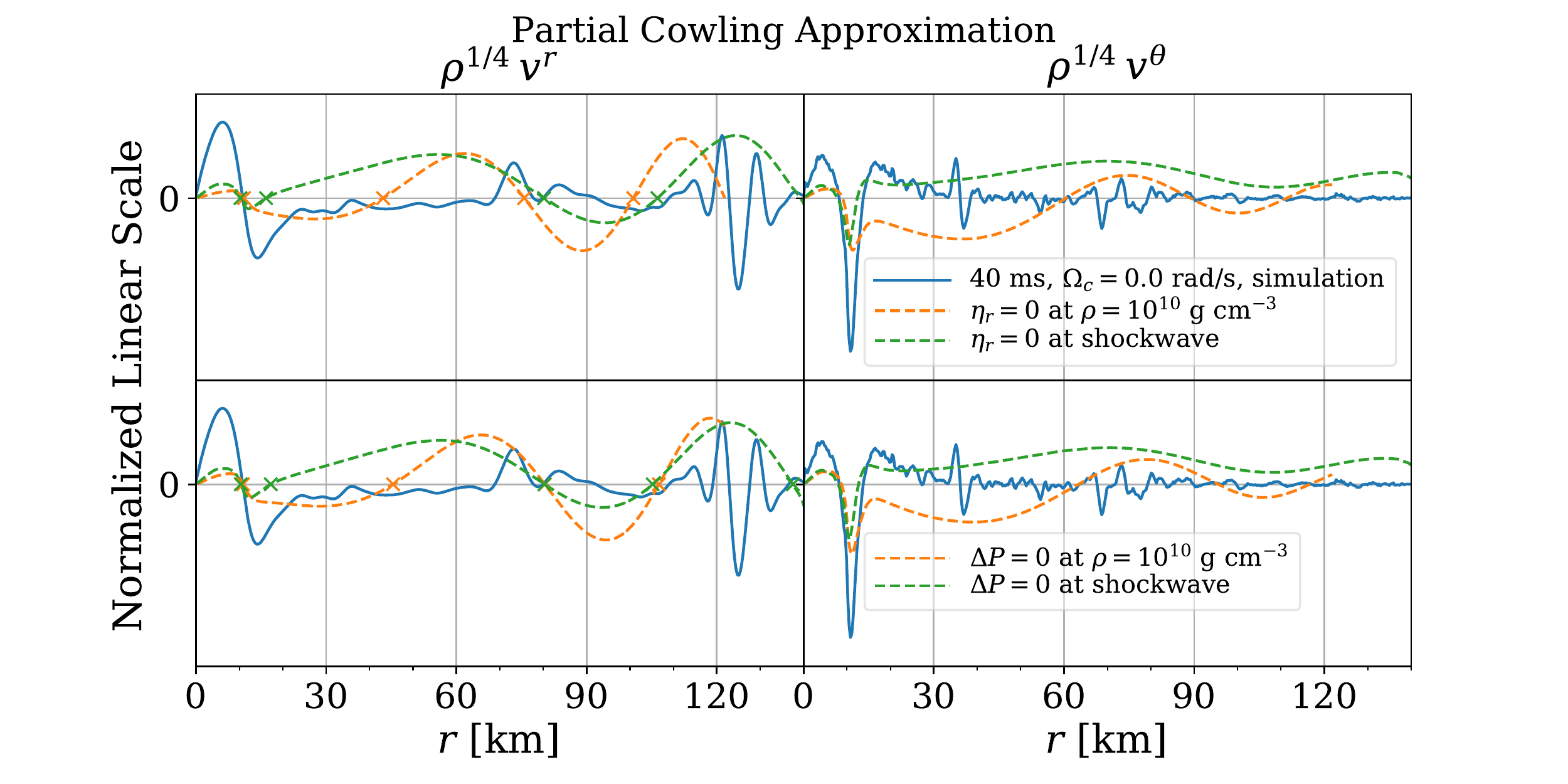}
\caption{Best-fit modefunctions from \fref{fig:fitquality_partialCowling}. Radial nodes are indicated with crosses.} \label{fig:modefunc_diffbcs_partialCowling}
\end{figure*}

In \fref{fig:fitquality_partialCowling} we plot the mismatch $\Delta$ for the modefunctions obtained using the partial Cowling approximation of~\cite{morozova2018gravitational}. When the outer boundary is placed at $\rho = 10^{10}\,$g$\,$cm$^{-3}$, the best-fit modefunction has a frequency closer to the simulation than when using the full Cowling approximation, although the quality of the modefunction fit is significantly worse. One can therefore be mislead by mode frequency matching to believe that the fit has improved when partially relaxing the Cowling approximation, when in fact it has become worse. The corresponding mode functions for the different boundary conditions are plotted in \fref{fig:modefunc_diffbcs_partialCowling}.

%
%
\section{Spectral filter kernels} \label{sec:kernels}
\begin{figure*}[htbp]
\centering
\includegraphics[width=\textwidth]{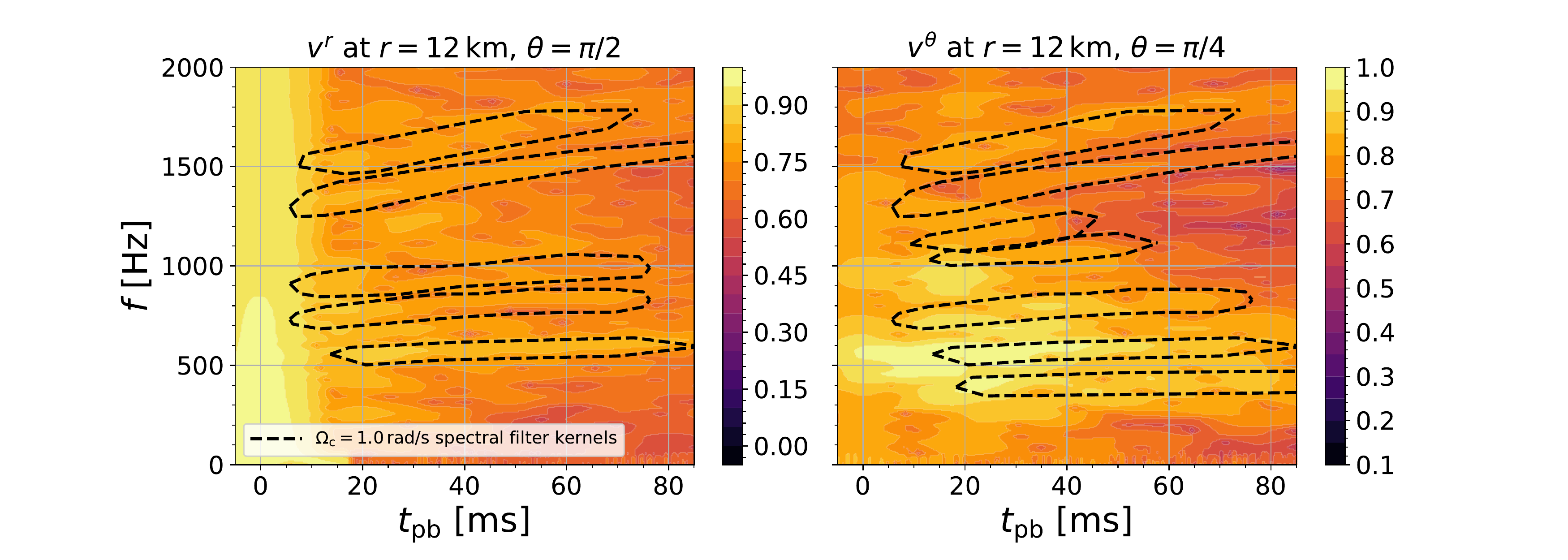}
\caption{A sampling of velocity spectrograms from the $\Omega_{\mathrm{c}}=1.0\,$rad/s simulation, with overlaid filtering kernels (dashed lines). The velocity spectrograms are normalized to 1 and displayed on a log$_{10}$ scale.} \label{fig:kernels}
\end{figure*}

For the interested reader, we plot the spectral filter kernel masks used in our modefunction matching analysis in \fref{fig:kernels}, on top of a sampling of velocity spectograms. In~\cite{westernacher2018turbulence}, shrinking these kernels in their frequency extent by a factor of 2 was found not to affect the mode function matching.

\renewcommand\bibname{References}

\bibliographystyle{apsrev4-1}
\bibliography{fluidbib}

\end{document}